\documentclass[12pt]{iopart}
\usepackage{cite,amssymb,amsfonts,graphicx,bm,dcolumn,mathrsfs}

\usepackage{epsfig}

\newcommand{\dd}{\mbox{d}}

\newcommand{\er}{\mathrm{e}}
\newcommand{\bea}{\begin{eqnarray}}
\newcommand{\eea}{\end{eqnarray}}
\newcommand{\be}{\begin{equation}}
\newcommand{\ee}{\end{equation}}

\begin{document}

\title{Ensemble inequivalence in Ising chains with competing interactions}
\author{Alessandro Campa$^1$\footnote{Corresponding author}, Vahan Hovhannisyan$^2$, Stefano Ruffo$^{3,4}$ and Andrea Trombettoni$^{5,6}$}
\address{$^1$ National Center for Radiation Protection and
Computational Physics, Istituto Superiore di Sanit\`{a},
Viale Regina Elena 299, 00161 Roma, Italy, and INFN Roma1, 00185 Roma, Italy}
\address{$^2$ A. I. Alikhanyan National Science Laboratory, 2 Alikhanyan Brothers Street, 0036 Yerevan, Armenia}
\address{$^3$ SISSA, via Bonomea 265, I-34136 Trieste, Italy \&
INFN, Sezione di Trieste, I-34151 Trieste, Italy}
\address{$^4$ Istituto dei Sistemi Complessi, CNR, via Madonna del Piano 10,
I-50019 Sesto Fiorentino, Italy}
\address{$^5$ Department of Physics, University of Trieste, Strada Costiera 11, I-34151 Trieste, Italy} 
\address{$^6$ CNR-IOM DEMOCRITOS Simulation Center, Via Bonomea 265, I-34136 Trieste, Italy}
\ead{\mailto{alessandro.campa@iss.it}, \mailto{v.hovhannisyan@yerphi.am}, \mailto{stefano.ruffo@gmail.com} and \mailto{andreatr@sissa.it}}

\begin{abstract}
We study the effect of competing interactions on ensemble inequivalence. We consider a one-dimensional Ising model with ferromagnetic mean-field interactions and short-range
nearest-neighbor and next-nearest-neighbor
couplings which can be either ferromagnetic or antiferromagnetic. Despite the relative simplicity of the model, our calculations in the microcanonical ensemble reveal a rich phase diagram. The comparison with the corresponding phase diagram in the canonical ensemble shows the presence of phase transition points and lines which are different in the two ensembles. As an example, in a region of the phase diagram where the canonical ensemble shows a critical point and a critical end point, the microcanonical ensemble has an additional critical point and also a triple point. The regions of ensemble inequivalence typically occur at lower temperatures and at larger absolute values of the competing couplings. The presence of two free parameters in the model allows us to obtain a
fourth-order critical point, which can be fully characterized by deriving its Landau normal form.
\end{abstract}

\maketitle

\section{Introduction}
\label{sec:intro}

The equivalence of thermodynamic behavior for a physical system across different ensembles is a cornerstone of statistical mechanics~\cite{Ruelle1969,Huang1987}. A major
example is provided by the equivalence of canonical and microcanonical ensembles, the former referring to macrostates of a system in thermal equilibrium with a heat bath at a fixed temperature and the latter to macrostates with a fixed total energy.

As a matter of fact, a detailed scrutiny of the proof of ensemble equivalence shows that it holds in the thermodynamic limit under very general assumptions for systems interacting via short-range interactions \cite{Ruelle1969}, when energy is additive. In the last decades, a considerable amount of attention has been devoted to establish whether ensemble equivalence holds when the constituents of a system do not interact by local interactions. In presence of a {\it long-range} interaction -- i.e., an
interaction between the microscopic constituents decaying with the distance $r$ more slowly than $r^{-d}$, where $d$ is the embedding dimension -- most often we witness
ensemble inequivalence~\cite{Campa2014}. The equilibrium state of the system, as defined by the value of the macroscopic observables, can depend on those that are taken as control variables. In particular, while in the microcanonical ensemble the value of the control variable, the energy, can assume any desired value, say $E^*$ (possibly within
a range allowed by the physical nature of the system), in the canonical ensemble, where the control variable is the temperature, it can happen that there is no temperature for which the average energy of the system is $E^*$.

In non-additive systems, like those in which interactions are long-range, there can be energy ranges where the microcanonical entropy is a non-concave function of the energy \cite{Campa2014}. This causes inequivalence, since the canonical entropy is always concave. More precisely, the canonical entropy is the concave envelope of the microcanonical one. The energies where the two entropies do not coincide are forbidden in the equilibrium states of the canonical ensemble and, in this ensemble, the system exhibits a first-order phase transition. Unlike additive systems, there is no phase separation with the coexistence of two phases in two subsystems.

These results clearly show that one can have ensemble inequivalence in presence of long-range interactions \cite{Campa2014,Defenu2023}. An arena in which these issues has been profitably investigated is provided by spin systems \cite{Campa2009}. A typical example is the Ising model with all-to-all couplings and a local nearest-neighbor (NN) interaction, a model which is solvable in dimension $d=1$ \cite{Baker1963,Nagle1970,Kardar1983}. As a result of the presence of both short- and long-range interactions, the thermodynamic and dynamical behavior of the system in the canonical and microcanonical ensembles may be different: one finds that the two ensembles have different phase
diagrams \cite{Mukamel2005}. Similar results hold for rotators in presence of both long- and short-range interactions
and in a variety of other spin models, see Refs. \cite{Campa2009,Campa2014}.

We like to mention also the relevance of the problem of ensemble equivalence in the study of systems made of a relatively small number of particles~\cite{book_Gross,droplets2}.
Indeed, in some cases canonical temperature can be well defined even for isolated small systems~\cite{Pethick2008,droplets}, which are formally described only in the microcanonical ensemble~\cite{book_Gross}. On the contrary, near a phase change of a small system, the two ensembles could give different predictions: this is the case of the experimental detection of negative specific heat~\cite{droplets2}.

Here, we want to address the question of how competing interactions affects ensemble inequivalence. Competing interactions  can be a source of complex behaviours and structures \cite{Chaikin1995}. When the two competing interactions act on a similar length scale, frustration can emerge~\cite{Mezard1987} since energy must be minimized respecting the constraints. If, in addition to competing interactions, there are also long-range interactions, the situation can become even more interesting.
In general, the presence of two forces that act on very different scales, one being much larger than the other, may result into the formation of patterns that grow from instabilities \cite{Seul1995}.
In particular, competing interactions have been widely studied in spin systems. In presence of competing ferromagnetic and/or antiferromagnetic couplings, a rich variety of ground states, stripe patterns and structures of multiple correlations and modulation lengths have been deeply investigated \cite{Diep2004}. First-order transitions are very likely to occur in the presence of competing interactions, for example when, in spin systems, there are both ferromagnetic interactions favoring aligned spins and antiferromagnetic interactions favoring anti-aligned spins.

In \cite{jpa2019} an Ising spin system with long-range interactions and two competing short-range interactions has been studied in the canonical ensemble. The very rich structure of the thermodynamic phase diagram exhibited by such a relatively simple spin system features $8$ distinct phase structures, with a variety of critical points. 
A peculiar feature of this model, which justifies the study of ensemble inequivalence, is the complexity of the ground state, exhibiting, besides the usual ferromagnetic and paramagnetic phases, also a ferrimagnetic one. We expect that this new feature creates novel patterns of ensemble inequivalence. In order to have three phases with only one order parameter, there must be two phases which are magnetized, and this is a consequence of the presence of different magnetized states already in the ground state.
In \cite{jsp2021}
the same model was used as a benchmark to illustrate a method to compute the microcanonical entropy at fixed magnetization without direct counting. 
Since long-range interactions may feature ensemble inequivalence and short-range couplings with opposite sign give rise to competing effects, the Ising model with an all-to-all long-range coupling and two short-range terms, is arguably the simplest model in which one can explicitly study the effect of competing interactions on ensemble inequivalence. 

In this paper, we extend the analysis by solving the model in the microcanonical ensemble. We discuss the properties of the microcanonical phase diagram and show that it is even richer than the canonical one, presenting some very unusual features. Our main motivation is two-fold. On one hand, we want to put in evidence ensemble inequivalence and the interesting structure of the microcanonical phase diagram. On the other hand, taking inspiration from the presence of a fourth-order critical point in the phase diagram (a concept that will be clarified in the following), we present an analysis of the general form of the entropy in the neighborhood of such point. This analysis aims at extending to singularities of higher order, at least for a particular case, the analogous analysis discussed in \cite{bb2005}.

We underline that the results presented in this paper will be focused on the characteristics associated with antiferromagnetic nearest-neighbor and
next-nearest-neighbor (NNN)couplings,
where the interesting structures are present, this being in common with the results that had been found for the canonical case \cite{jpa2019}. This means that the most interesting and complex structures in the phase diagram occur where the competition among the interactions is strongest; in fact, with antiferromagnetic nearest-neighbor
and next-nearest-neighbor interactions, these interactions compete among them and both compete with the mean-field ferromagnetic interaction.

We would like to anticipate a particular feature that has been found. Our analysis has unveiled the presence, for a range of the NNN coupling constant, of a quite small complex structure, with a critical point, a triple point and a critical end point, all very close to each other. The triple point is no more present when we consider a larger range of this coupling, but we have found that a structure with a critical point and a critical end point, which are extremely close, persists for a wide range of the NNN coupling. Details on this finding are given in the section where we present the phase diagrams.

The plan of the paper is the following: In Section \ref{sec:the_model} we introduce the model studied in the rest the paper and remind its properties in the canonical ensemble. 
In Section \ref{solandres} we present the solution of the model in the microcanonical ensemble, with the microcanonical entropy determined in \ref{microen} and properties of the microcanonical phase diagram discussed in \ref{microphase}. A study of the normal form for a fourth-order critical point is presented in Section \ref{normalform}, and our conclusions are in Section \ref{sec:conclusions}. Appendices present additional material with details of the calculation presented in the main text. 

\section{The model}
\label{sec:the_model}

We consider a one-dimensional (1D) lattice, where in each one of the $N$ sites there is an Ising spin variable $S_i$ with two possible values, $+1$ and $-1$. The interactions between the spins are given by: an all–to–all mean–ﬁeld ferromagnetic coupling, a coupling between NN spins and a coupling between
NNN spins.
We denote by $J$ the mean–ﬁeld long–range coupling and by $K_1$ and $K_2$ the (ferromagnetic or antiferromagnetic) NN and NNN couplings, respectively. 
Then, the Hamiltonian has the form:
\begin{equation}
H = - \frac{J}{2N} \sum_{i,j} S_i S_j - \frac{K_1}{2} \sum_i S_i S_{i+1} -
\frac{K_2}{2} \sum_i S_i S_{i+2} \,.
\label{H}
\end{equation}
We assume periodic boundary conditions. We consider $J > 0$ and, without loss of generality, we can take $J = 1$ (this is equivalent to measuring the energy and temperature in units of $J$ and substituting
$K_i$ with $K_i/J$). Notice that if $J<0$ there is no order at finite temperature.

Depending on the sign of the other parameters, $K_1$ and $K_2$, we can have competing
interactions. 
The mean–ﬁeld ferromagnetic interaction ($J>0$) favours aligned spins. A negative value of $K_1$ corresponds to an antiferromagnetic coupling between NN sites and would prefer NN spins with opposite alignments. When both $K_1$ and $K_2$ are negative there is competition, since a negative $K_1$ prefers
alternating spins, a conﬁguration where NNN are aligned, which is not favoured by a negative value of the coupling constant $K_2$ between NNN spins. These competitions give rise to a very rich phase diagram, as discussed below. 

When $K_2 = 0$, the model has been solved both in the canonical and microcanonical ensembles \cite{Kardar1983,Mukamel2005}. In the canonical ensemble, by performing a Hubbard-Stratonovich transformation, one can write the partition function as
\begin{equation}
Z=\sqrt{\frac{\beta N}{2 \pi}} \int_{-\infty}^{+\infty} dx \, \er^{-N \beta \left( \frac{x^2}{2}+f_0\right)}\,.
\label{Z_K1_0}
\end{equation}
In Eq. (\ref{Z_K1_0}) $x$ is the variable introduced in the Hubbard-Stratonovich transformation \cite{Kardar1983_bis}, $\beta=1/T$ (where $T$ is the temperature and $k_B=1$) and $f_0$ is the free energy density of the 1D classical NN Ising model with coupling $K_1/2$ in a magnetic field $x$ \cite{Huang1987}. In the  $(K_1,T)$ phase diagram there is a line of second-order phase transitions
with mean-field critical exponent, defined by $\beta_c = \er^{-\beta_c K_1}$.
This line ends at a tricritical point defined by the equation $\beta_{TP} K_{1;TP} = -(1/2) \ln 3$.
Further lowering the temperature one has a first-order phase transition line reaching $T=0$ for $K_1 = - 1/2$. In the microcanonical ensemble, 
the model, for $K_2=0$, exhibits a critical line characterized by the same expression found in the canonical ensemble.
However, this line extends beyond the canonical tricritical point, reaching a microcanonical tricritical point $K_{1;MTP}$. For $K_1 < K_{1;MTP}$ 
the transition becomes first-order and has a discontinuity in the temperature. 

Let us now consider the case $K_2 \neq 0$. In the canonical ensemble, this model has been solved in  \cite{jpa2019}, again using a suitable Hubbard-Stratonovich transformation. It is convenient to plot the phase diagram in the $(K_1,T)$ plane at fixed values of $K_2$. One finds eight regions of the values of $K_2$ with qualitatively distinct properties, each one characterized with different first- and second-order phase transition lines and critical points. One introduces the magnetization, the NN correlation and the NNN correlation respectively as $m=(1/N) \sum_i S_i$, $g_1 = (1/N) \sum_i S_i S_{i+1}$ and $g_2 = (1/N) \sum_i S_i S_{i+2}$. The canonical phase diagram structure of the
$K_2=0$ case qualitatively persists for all positive values of $K_2$ and for negative values of $K_2$ down to $K_2 \simeq -0.0885$. For $K_2$ between
$\simeq -0.1542$ and $\simeq -0.0885$ below the second-order transition line a first-order line emerges. 
When $K_2$ is between $-1/6$ and $\simeq - 0.1542$ the first-order line bifurcates at a triple point and one of the two first-order line terminates at a new tricritical point where it meets the second-order line. At $K_2=-1/6$ the triple point gives rise to two first-order lines by further decreasing $K_2$ till $K_2 \simeq - 0.2672$. In the region with $K_2$ between $\simeq - 0.2745$ and $\simeq - 0.2672$ the second-order line of the previous region breaks in two pieces and a first-order line, limited by two tricritical points, so to have three first-order lines. Two first-order lines merge at $K_2 \simeq -0.2745$ and a second-order line connects to this merged first-order line, with the latter ending
in a tricritical point and having a critical end point with $K_2$ down to $K_2 \simeq - 0.2953$. A first-order line bifurcates and a triple point emerges at
$K_2 \simeq -0.2953$ and in the region with $K_2$ between $\simeq -1/3$ and $\simeq -0.2953$ two first-order lines ends at two tricritical points separating them from two second-order lines. At $K_2=-1/3$ a first-order line no longer separates two regions with non-vanishing magnetization, and the region
with zero magnetization persists at very low temperature down to $T=0$. For $K_2 < -1/3$ this part of the phase diagram is surrounded by two first-order lines ending in tricritical points,
from which two second-order lines depart. We refer to \cite{jpa2019} for a detailed discussion of the eight regions of values of $K_2$ just described.

\section{Solution of the model in the microcanonical ensemble} 
\label{solandres}
\subsection{The microcanonical entropy}
\label{microen}
The solution of the model in the microcanonical ensemble requires the computation of the entropy in that ensemble. The procedure would be to compute the number
of states for any given value of the energy $E$. However, as we have shown in a previous work \cite{jsp2021}, already for a relatively simple system like the one we
are studying, this direct counting technique is very cumbersome, and a faster way is the so called min-max procedure \cite{Campa2009}.
This can be adopted in systems where the canonical partition function of a system of $N$ particles can be expressed as
\be
\label{hstrans}
Z = C \int_{-\infty}^{+\infty} \dd x \, \er^{- N\beta \tilde{f}(\beta,x)} \, ,
\ee
where $\beta=1/T$ is the inverse temperature, and $x$ is an auxiliary variable. In the thermodynamic limit the value of this integral is approximated by the maximum
of its integrand (meaning that the corrections are subleading in $N$), so that the canonical free energy per particle, $f(\beta)$, is given by
\be
\label{freehs}
\beta f(\beta) = \min_x [\beta\tilde{f}(\beta,x)] \, .
\ee
An expression like (\ref{hstrans}) can be obtained in our case through the Hubbard-Stratonovich transformation \cite{jpa2019}
\be
\label{hsham}
\exp \left[ \frac{\beta}{2N} \left(\sum_{i=1}^N S_i\right)^2\right] = \sqrt{\frac{\beta N}{2\pi}} \int_{-\infty}^{+\infty} \dd x \,
\exp \left[ -\frac{N\beta}{2}x^2 + \beta x \sum_{i=1}^N S_i \right] \, .
\ee
Before giving the concrete expression of $\tilde{f}(\beta,x)$, computed in Ref. \cite{jpa2019}, we underline that the min-max
procedure allows to compute the microcanonical entropy, as a function of the energy per particle $\epsilon = E/N$, as \cite{Campa2009}
\be
\label{minmaxmic}
s(\epsilon) = \max_x \left\{ \min_\beta \left[ \beta \epsilon - \beta \tilde{f}(\beta,x) \right] \right\} \, .
\ee
As remarked above, in non-additive systems the function $s(\epsilon)$ is not always concave, and in the energy ranges where it is not concave
we have ensemble inequivalence. In the canonical ensemble the entropy is given by
\be
\label{minmaxcan}
s_{\rm can}(\epsilon) = \min_\beta \left\{ \max_x \left[ \beta \epsilon - \beta \tilde{f}(\beta,x) \right] \right\} \, ,
\ee
which is always concave; this is assured by the properties of the Legendre-Fenchel transform. We remind that the Legendre-Fenchel
transform of a function $f(x)$ is a function $g(y)$ defined by
\be
\label{deflegfen}
g(y)=\min_x [yx-f(x)]
\ee
(a mathematically more rigorous definition would employ the infimum instead of the minimum). This transform is very important in the theory of
equivalence and nonequivalence of ensembles. A detailed study of this subject can be found in Ref. \cite{Touch2015}. Here we just mention
the relevant fact that, in Eq. (\ref{deflegfen}), $g(y)$ is concave regardless of the form of $f(x)$. Then we see that $s_{\rm can}(\epsilon)$ given in
Eq. (\ref{minmaxcan}) is concave. On the other hand, the function $s(\epsilon)$ of Eq. (\ref{minmaxmic}) is not a Legendre-Fenchel transform, and it
can be non-concave.

The function $\tilde{f}(\beta,x)$ is given by
\be
\label{ftilde}
\tilde{f}(\beta,x) = \frac{1}{2}x^2 -\frac{1}{2\beta}\ln \lambda (\beta,x) \, ,
\ee
where $\lambda(\beta,x)$ is the largest eigenvalue of the transfer matrix \cite{jpa2019}
\be
\label{transmat}
T =  \left(
\begin{array}{cccc}
e^{ \beta \left( K_1 + K_2 + 2x \right)} &  e^{ \beta \left( \frac{K_1}{2} + x \right)} & e^{ \beta \left( -\frac{K_1}{2} + x \right)} &
e^{ -\beta K_2}  \\
e^{ \beta \left( -\frac{K_1}{2} + x \right)} & e^{ \beta \left( - K_1 + K_2 \right)} &  e^{ -\beta K_2} &
e^{ \beta \left( \frac{K_1}{2} - x \right)} \\
e^{ \beta \left( \frac{K_1}{2} + x \right)} &  e^{ -\beta K_2} & e^{ \beta \left( - K_1 + K_2 \right)} &
e^{ \beta \left( -\frac{K_1}{2} - x \right)} \\
e^{ -\beta K_2} & e^{ \beta \left( -\frac{K_1}{2} - x \right)} & e^{ \beta \left( \frac{K_1}{2} - x \right)} &
e^{ \beta \left( K_1 + K_2 - 2x \right)}
\end{array} \right).
\label{Talt}
\ee
Although this matrix is not symmetric and can have complex eigenvalues, its elements are all strictly positive, and the Perron-Frobenius
theorem \cite{Perron1907,Frobenius1912} guarantees that the eigenvalue with the largest modulus is real and positive.

In the following we show the results that we have obtained for the microcanonical phase diagram. To highlight explicitly the ensemble inequivalence, in all cases (except the last one) we
have shown also the canonical phase diagram. As already remarked, the latter was analyzed in Ref. \cite{jpa2019}.

\subsection{The microcanonical phase diagram}
\label{microphase}

\begin{figure}[htbp]
\begin{center}
\begin{tabular}{cc}
 \includegraphics[clip, trim=3.5cm 8.5cm 3.5cm 8.5cm, width=0.49\textwidth]{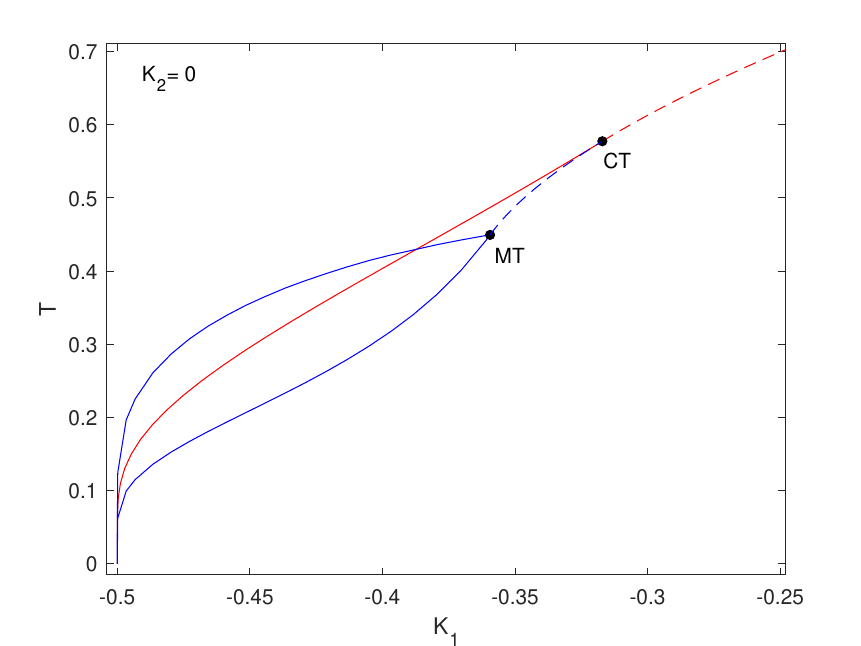} &
 \includegraphics[clip, trim=3.5cm 8.5cm 3.5cm 8.5cm,  width=0.49\textwidth]{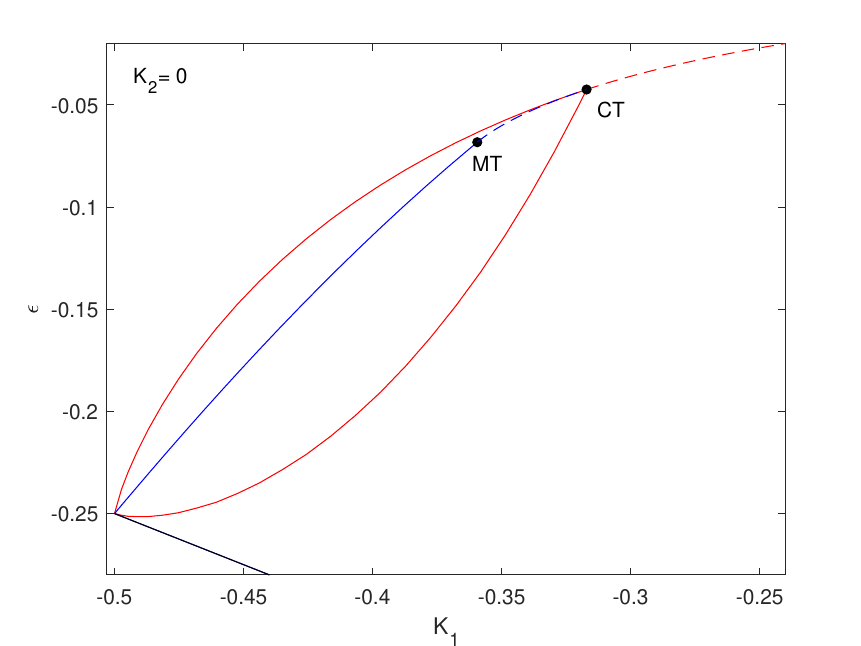}
\end{tabular}
\caption {The $(K_1, T)$ (left) and $(K_1,\epsilon)$ (right) phase diagrams of the canonical and microcanonical ensembles at $K_2=0$. The dashed red and blue lines denote the lines of second-order transitions, ending at the canonical tricritical point (CT) and at the microcanonical tricritical point (MT) for the canonical and microcanonical ensembles, respectively. Red (blue) solid lines are the first-order transition lines for the canonical (microcanonical) ensemble. In the $(K_1, \epsilon)$ phase diagram the black solid line denotes the $K_1$-dependent ground state energy. The same representation for the different lines is adopted in all the following figures of phase diagrams.}
\label{fig1}
\end{center}
\end{figure}

In this section we show the complete analysis of ensemble inequivalence. It is important to note that phase diagrams plotted here are 2D cuts of the 3D phase diagram, the cut being defined by the choice of a fixed value of $K_2$. To achieve a comprehensive understanding of the influence of the next-nearest-neighbor interaction on the phase diagram, we have started our analysis by recalling the phase diagram corresponding to $K_2=0$; as already mentioned, the model with $K_2=0$ has already been solved both in the
canonical \cite{Nagle1970,Kardar1983} and in the microcanonical case \cite{Mukamel2005}. In figure \ref{fig1} we have plotted the $(K_1, T)$ (left) and the ($K_1, \epsilon)$ (right) phase diagrams for a value of $K_2=0$. In these plots, dashed lines represent second-order transitions, while solid lines indicate first-order transitions. The lines of the canonical ensemble are in red, while those of the microcanonical ensemble are in blue. The same representation is adopted also in all the other phase diagrams, and we will not repeat this specification. We have chosen to display the most interesting parts of the phase diagrams; at values of $K_1$ larger than those plotted, we have simply the indefinite continuation of the second-order transition line. We remark that, while the phase diagrams with the temperature $T$ as one of the coordinates are the ones most commonly presented, the phase diagrams with, instead, the energy $\epsilon$ taken as the thermodynamic variable, seem to be more natural for the microcanonical case, being the energy the thermodynamic control variable in this ensemble. Besides, the study of the phase diagrams in both versions facilitates a comparison between the two ensembles. As usual, the microcanonical second-order line (also called critical line, we will use interchangeably the two wordings) extends farther compared to the canonical one \cite{Campa2009}, and the second-order phase transition line coincides in both ensembles up to the canonical tricritical (CT) point. After this point the canonical ensemble has a first-order transition line (the red line), while for the microcanonical ensemble the critical line continues up to the microcanonical tricritical (MT) point. The microcanonical first-order transitions are characterized by a temperature jump \cite{Campa2009}, therefore the $(K_1,T)$ diagram has two branches in correspondence to the microcanonical first-order transitions. The first-order canonical and microcanonical transition lines meet at the $T = 0$ axis, at the value of $K_1$ equal to $-0.5$. In the $(K_1,\epsilon)$ phase diagram it is the canonical ensemble that exhibits two branches associated with first-order transitions: each branch corresponds to the energy value of one of the two extremes of the energy jump occurring in the transition and usually represented, in the caloric curves $(\epsilon,T)$, with the associated Maxwell construction. As a consequence, the region between the two red solid lines in the $(K_1,\epsilon)$ diagram is forbidden in the canonical ensemble \footnote{The analogous region between two solid blue line in the $(K_1,T)$ diagram cannot be considered forbidden in the microcanonical ensemble. In fact, since the temperature jump in a microcanonical first-order transition is negative, the systems obtains before and after the transition the temperatures in the range characterizing the jump.}. In the $(K_1,\epsilon)$ diagram (here as in the following), a black solid line marks (for the values of $K_1$ where it is included in the energy range shown in the plot) the $K_1$-dependent ground energy of the system; there are no states below this line. In this diagram the first-order canonical and microcanonical transition lines meet at the ground energy.

We conclude the presentation of the phase diagrams for $K_2=0$ by writing explicitly the values of the coordinates of the relevant points. They are: 
\begin{itemize}
  \item canonical tricritical (CT) point: $K_{1} \backsimeq -0.3171, T  \backsimeq 0.5775, \epsilon  \backsimeq -0.04246$.
  \item microcanonical tricritical (MT) point: $K_{1} \backsimeq -0.3594, T  \backsimeq 0.4495, \epsilon \backsimeq -0.06825$.
 \end{itemize}
 We will do the same for all the other values of $K_2$ that will be presented. In these lists of the relevant points it is possible to see the acronyms used for them in the figures.

\begin{figure}[htbp]
\begin{center}
\begin{tabular}{cc}
 \includegraphics[clip, trim=3.5cm 8.5cm 3.5cm 8.5cm, width=0.49\textwidth]{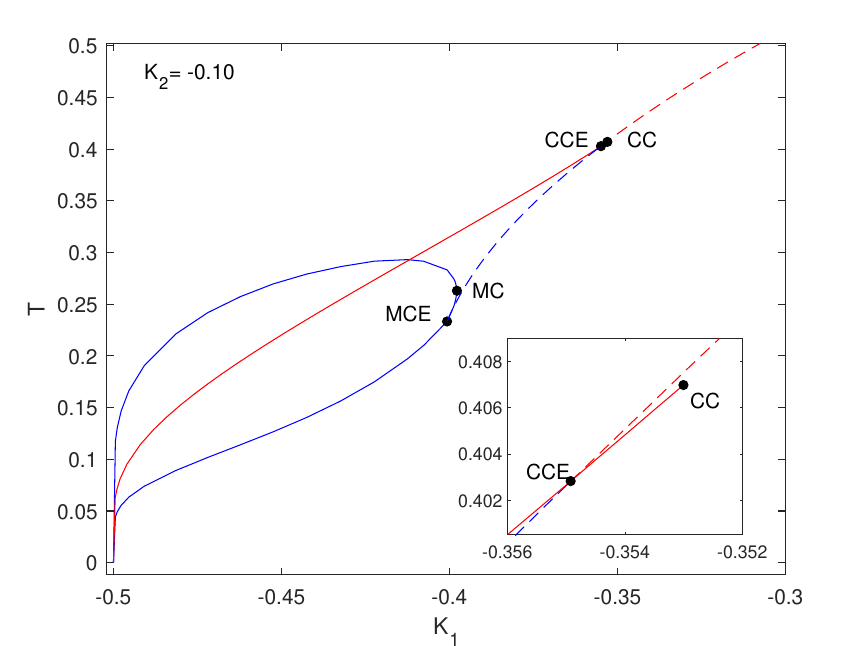} &
 \includegraphics[clip, trim=3.5cm 8.5cm 3.5cm 8.5cm,  width=0.49\textwidth]{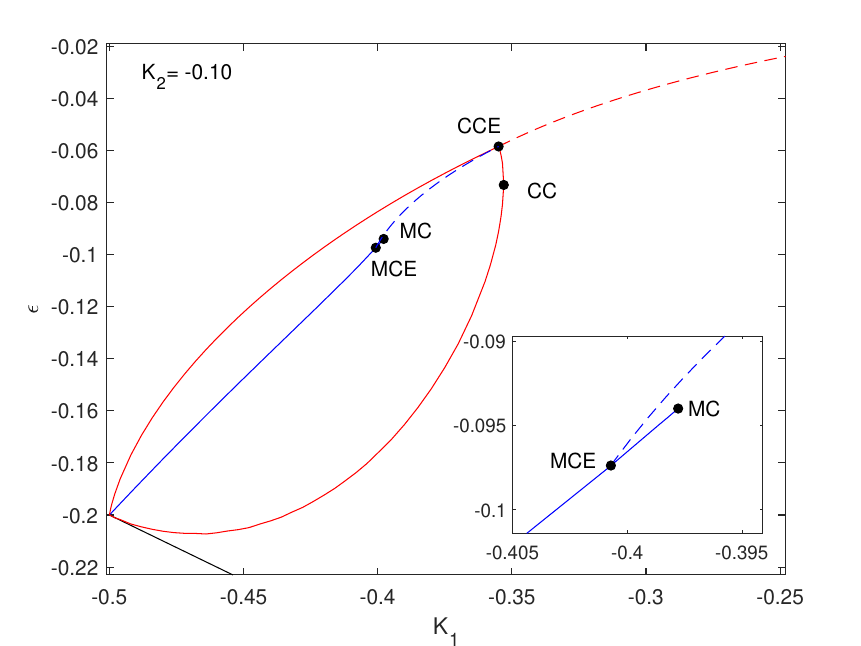}
\end{tabular}
\caption {The $(K_1, T)$ (left) and $(K_1,\epsilon)$ (right) phase diagrams of the canonical and microcanonical ensembles at $K_2=-0.10$. The lines of second-order transitions now end, for both ensembles, at a critical end point, marked with CCE and MCE for the canonical and microcanonical ensembles, respectively, where they meet the first-order transition lines. The latter start at $T=0$ (at the ground energy) in the $(K_1,T)$ diagram (in the $(K_1,\epsilon)$ diagram), and after the critical end point they continue up to a critical point, CC and MC for the canonical and microcanonical ensemble, respectively. The insets show zoomed regions around close relevant points.}
\label{fig2}
\end{center}
\end{figure}

The $(K_1,T)$ and $(K_1,\epsilon)$ phase diagrams for the value $K_2=-0.10$ are reported in figure \ref{fig2}, in the left and right plots, respectively. Here again the microcanonical critical line extends farther with respect to the canonical one; however, the structure of the phase diagrams has changed with respect to those for $K_2=0$. In fact, the canonical critical line ends where it meets a first-order transition line at a canonical critical end point (CCE). The same occurs for the microcanonical critical line; it coincides with the canonical one up to the CCE, but it continues up to a microcanonical critical end point (MCE), where it meets a first-order transition line. For both ensembles, the first-order transition lines start at the same point at $K_1 = -0.5$ and $T=0$ (or the ground energy), and after the critical end point they continue up to a canonical (CC) or a microcanonical (MC) critical point. The section of the first-order line between the critical end point and the critical point, in both ensembles, separates two different magnetized states. This section is quite short. The reason is that
the structural change of the $(K_1,T)$ and $(K_1,\epsilon)$ phase diagrams, varying $K_2$ from $K_2=0$ to $K_2=-0.10$, has just occurred; in this change what was a tricritical point
has splitted in a critical end point and a critical point, which are still quite close. We will analyze in detail this situation in the following section, by showing how it can be
characterized with a Landau-type normal form for the expressions in the right hand side of Eqs. (\ref{freehs}) or (\ref{minmaxmic}). To have a better visualization of the section
of the first-order lines between the critical end point and the critical point, in the left plot of figure \ref{fig2} an inset displays a zoomed region around the CCE and the CC, while an analogous inset in the right plot of the figure shows a zoomed region around the MCE and the MC. It can be noted how in both cases the section of the first-order line between the critical end point and the critical point is very close to the second-order line.      

The coordinates of the relevant points of the phase diagrams are:
\begin{itemize}
  \item canonical critical (CC) point: $K_{1} \backsimeq -0.3530, T  \backsimeq 0.4070, \epsilon  \backsimeq -0.07327$.
  \item canonical critical end (CCE) point: $K_{1} \backsimeq -0.3549, T  \backsimeq 0.4028, \epsilon  \backsimeq -0.05849$.
  \item microcanonical critical (MC) point: $K_{1} \backsimeq -0.3978, T  \backsimeq 0.2630, \epsilon  \backsimeq -0.09400$.
  \item microcanonical critical end (MCE)  point: $\!\! K_{1} \backsimeq -0.4007, T  \backsimeq 0.2333, \epsilon  \backsimeq -0.09738$.
 \end{itemize}

\begin{figure}[htbp]
\begin{center}
\begin{tabular}{cc}
 \includegraphics[clip, trim=3.5cm 8.5cm 3.5cm 8.5cm, width=0.49\textwidth]{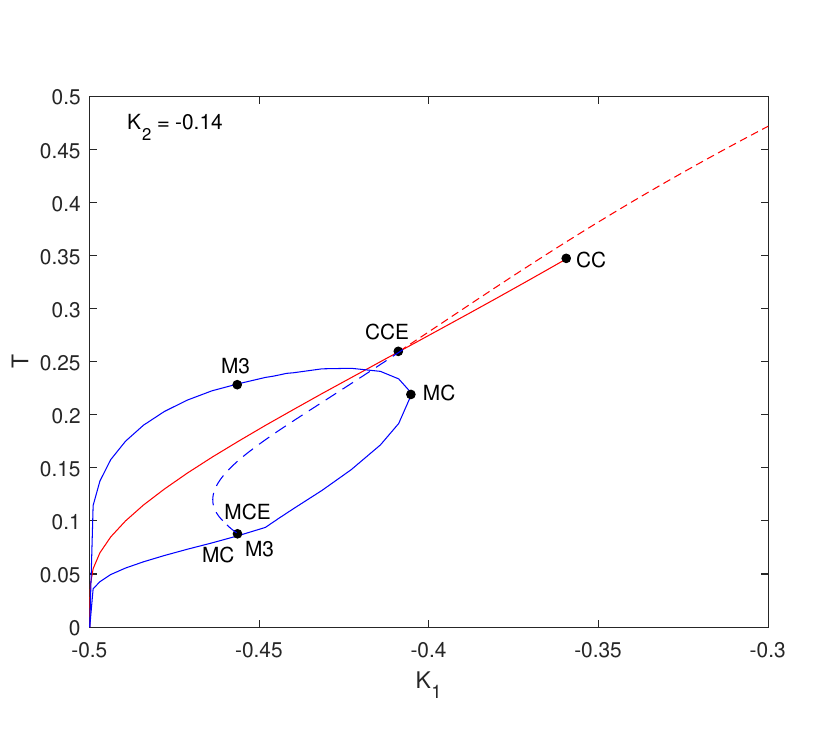} &
 \includegraphics[clip, trim=3.5cm 8.5cm 3.5cm 8.5cm, width=0.49\textwidth]{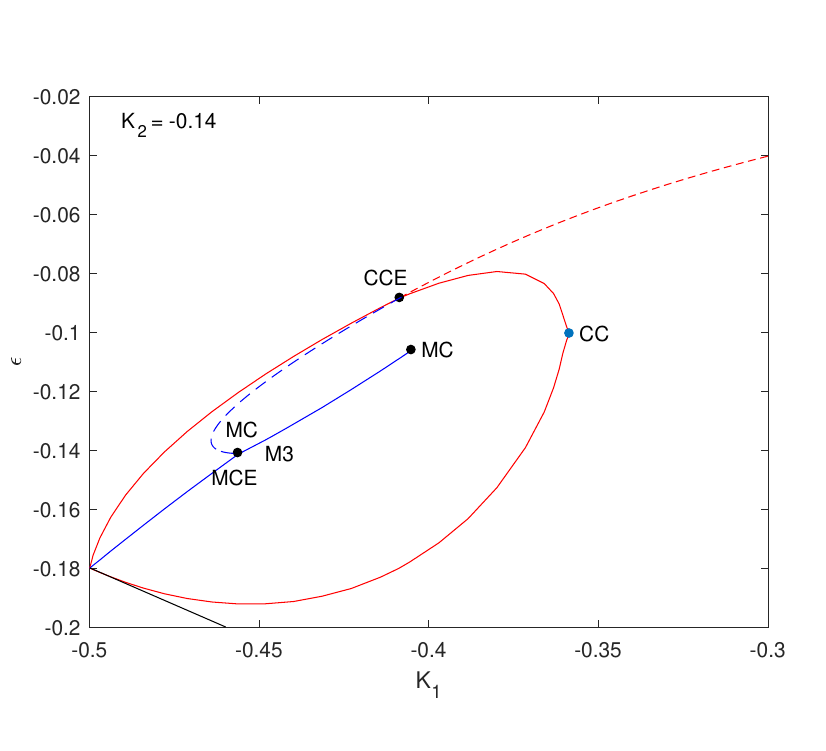}
\end{tabular}
\caption {The $(K_1,T)$ (left) and $(K_1,\epsilon)$ (right) phase diagrams of the canonical and microcanonical ensembles at $K_2=-0.14$. Like in the previous case, the
second-order lines end at a critical end point (CCE and MCE for the canonical and microcanonical case, respectively). However, while the structure of the canonical
diagram is similar to the previous case, the microcanonical diagram now presents a new and peculiar structure, with the appearance of another critical point (MC) and
a triple point (M3) very close to the MCE. To distinguish these three points a zoomed plot of that small region is shown in figure \ref{fig4}.}
\label{fig3}
\end{center}
\end{figure}

\begin{figure}[htbp]
\begin{center}
\begin{tabular}{cc}
 \includegraphics[clip, trim=3.7cm 7.7cm 3.7cm 8.5cm,  width=0.5\textwidth]{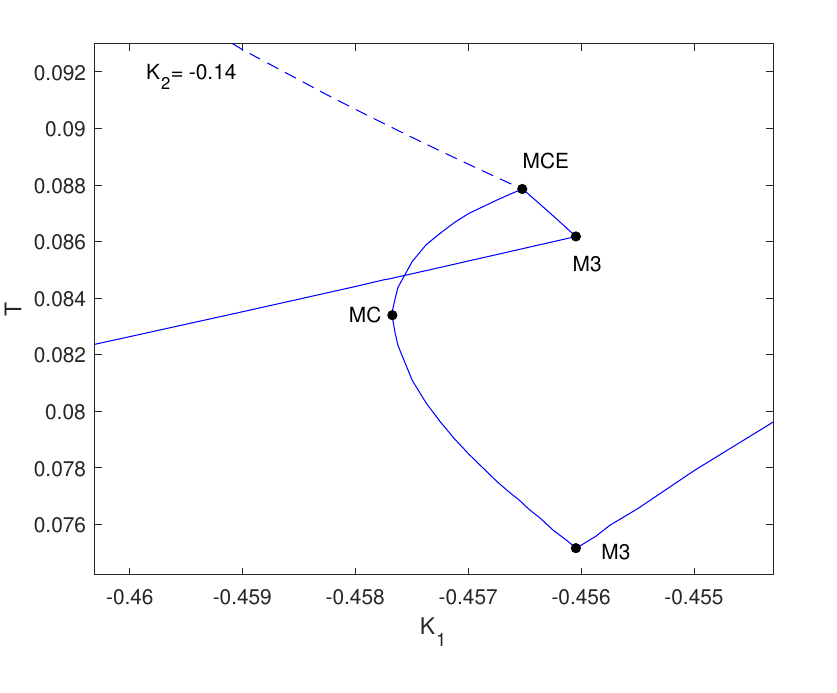} &
 \includegraphics[clip, trim=3.6cm 7.6cm 3.6cm 8.5cm,  width=0.5\textwidth]{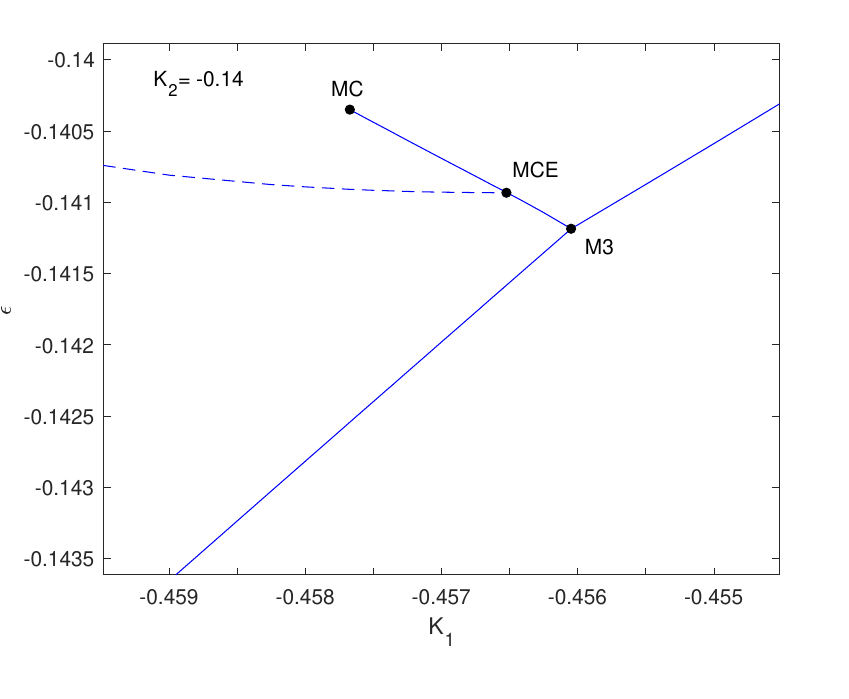}
\end{tabular}
\caption {The zoomed plots of the small region of the microcanonical $(K_1,T)$ (left) and $(K_1, \epsilon)$ (right) phase diagrams, at $K_2=-0.14$, around the MCE.
In the $(K_1,\epsilon)$ diagram a very short further first-order line, indistinguishable at the scale of the full diagram in the previous figure, appears. Associated to
this first-order transition there is, in the $(K_1,T)$ diagram, a looped structure, located, comparing with the $(K_1,T)$ full diagram in the previous figure, inside a
larger loop.}
\label{fig4}
\end{center}
\end{figure}

We now go to the phase diagrams for $K_2=-0.14$. As we will see in a moment, the microcanonical phase diagram in this case is very peculiar, and we have chosen to show four different plots of the phase diagrams: the two plots in figure \ref{fig3} concern the $(K_1,T)$ (left panel) and the $(K_1,\epsilon)$ (right panel) phase diagrams, shown in their entirety,
while in figure \ref{fig4} we show a particular zoomed region of the two diagrams, again in the left panel for the $(K_1,T)$ case and in the right panel for the $(K_1,\epsilon)$ case. The change of structure with respect to the previous case, $K_2=-0.10$, is related to the microcanonical phase diagram, while the canonical structure is the same of the previous case (obviously with different coordinates for the relevant points). The microcanonical structure is characterized,
among the other things, by the appearance of a microcanonical triple point (M3), but there are several interesting features that deserve detailed comments. Let us begin with the two full phase diagrams in figure \ref{fig3}. Comparing the left plot of figure \ref{fig3} with the left plot of figure \ref{fig2} and the right plot of figure \ref{fig3} with the right plot of figure \ref{fig2} we see from the red curves that the canonical phase diagram has, as remarked, the same qualitative structure; we only note that now the separation between the critical end point and the critical point is sufficient to distinguish them clearly in the full diagram (figure \ref{fig3} left). On the other hand, in the microcanonical phase diagram a complicated structure has appeared in a quite small region. In fact, in the two panels of figure \ref{fig3} we have denoted what appears to be a single point with MCE (microcanonical critical end point), MC (microcanonical critical point), and M3 (microcanonical triple point). To distinguish those three relevant points we need to show zoomed plots of that small region. Actually, since in the $(K_1,T)$ microcanonical diagram a triple point is characterized by three values of the temperature, what cannot be distinguished at the scale of figure \ref{fig3} are two of those temperatures, while the third one is quite different from those, and shown by the M3 mark in the upper part of the blue loop. A feature that is already clear in the full diagrams is that now there is a reentrant region in the microcanonical ensemble, caused by the turning of the critical line: for a range of $K_1$ values by increasing the energy $\epsilon$ one passes, with two continuous transitions, from an unmagnetized state to a magnetized state and again to an unmagnetized state (see also the related comment in the next paragraph).

\begin{figure}[htbp]
\begin{center}
\begin{tabular}{cc}
\includegraphics[clip, trim=0.7cm 9.7cm 3.7cm 8.5cm,  width=0.5\textwidth]{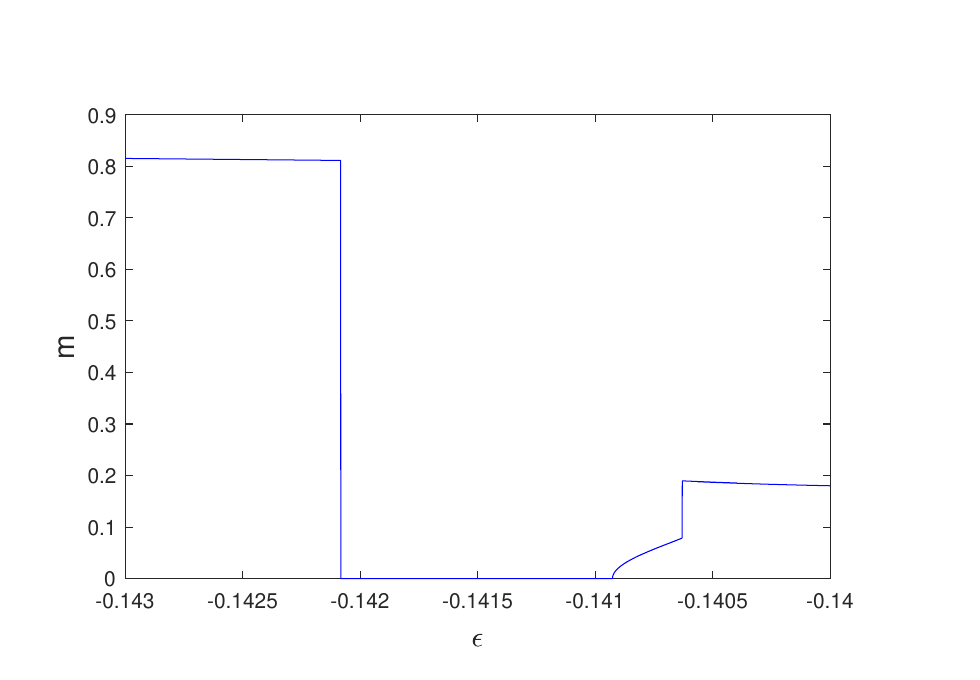} &
\includegraphics[clip, trim=0.7cm 9.6cm 3.6cm 8.5cm,  width=0.5\textwidth]{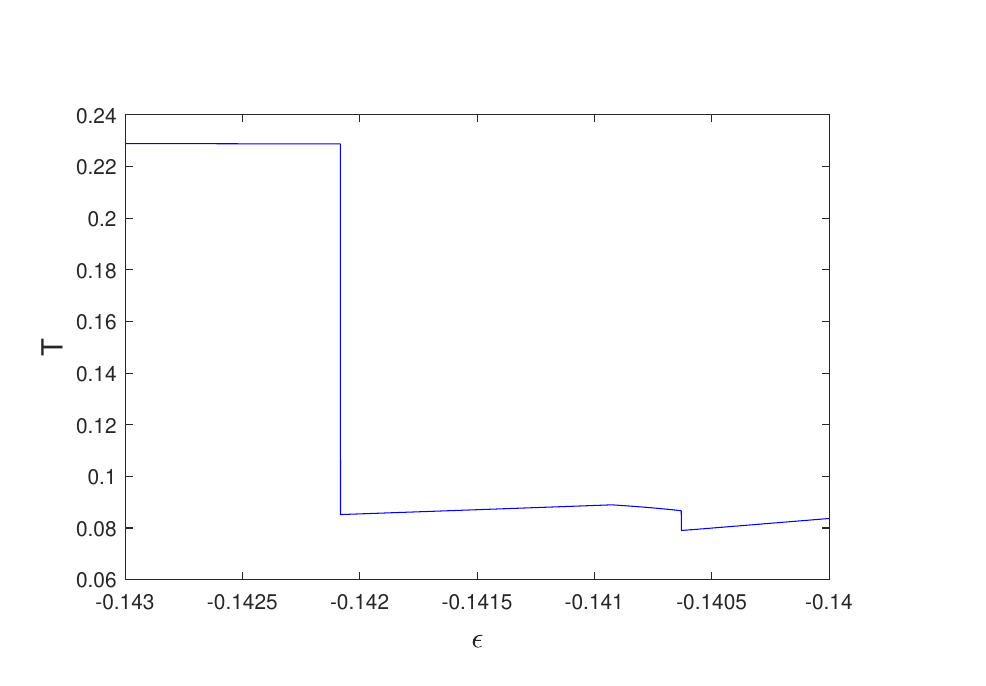}
\end{tabular}
\caption {The magnetization as a function of energy (left) and the caloric curve (right) for $K_2 = - 0.14$ and
$K_1 = - 0.457125$.}
\label{fig5}
\end{center}
\end{figure}

We now comment the zoomed plots in figure \ref{fig4}. In the small region represented in these plots there are only transition lines of the microcanonical ensemble. As a matter of fact we see, from figure \ref{fig3}, that in the $(K_1,\epsilon)$ diagram the small region is entirely between the two branches of the canonical first-order transition, and therefore it is forbidden in the canonical ensemble. In the plots of figure \ref{fig4} we see the final part of the second-order line, that ends at the microcanonical critical end point (MCE) (left panel for the $(K_1,T)$ diagram and right panel for the $(K_1,\epsilon)$ diagram). Let us begin the analysis with the $(K_1,\epsilon)$ diagram. In this plot we see that from the first-order line starting at the ground energy, and ending with the critical point in the right panel of figure \ref{fig3}, another short first-order line starts at the triple point M3; this short line, ending at another critical point MC, is not visible at the scale of figure \ref{fig3}, but only here in figure \ref{fig4}. We note that the critical end point MCE is located on this short first-order line. This line marks a transition between either a magnetized state and an unmagnetized state for energies lower than that of the MCE, or between two magnetized states for energies higher than that of the MCE. The energy of the triple point M3 is the common energy, at the transition point for that particular value of $K_1$, of these three states, the unmagnetized one and the two magnetized ones. To better visualize and explain what occurs we present, in figure \ref{fig5}, the microcanonical magnetization curve ($m$ vs. $\epsilon$) and the microcanonical caloric curve ($T$ vs. $\epsilon$); this for a value of $K_1$ equal to $-0.457125$, which is between that of MC and that of MCE.
In these two plots the lower energy value, $\epsilon = -0.143$, is below the lower first-order line in figure \ref{fig4} (right), while the upper value, $\epsilon = -0.140$, is above both first-order lines in that figure. Thus, we see from figure \ref{fig5} that at increasing energy within this range there is a first-order transition between a magnetized state to an unmagnetized state, with a corresponding negative temperature jump, followed by a second-order transition to a magnetized state, accompanied by a small kink in the caloric curve, then followed by a first-order transition between two magnetized states, associated by a small temperature jump. Referring to the left plot in figure \ref{fig4}, we conclude that soon after the former first-order transition the temperature increases from the value in the lower branch of the small loop to the value on the dashed line, then, after the second-order transition it decreases to the value in the upper part of the small loop, and finally at the latter first-order transition it jumps to the value in the lower full line. Not represented in the plots of figure \ref{fig5}, at higher energies, around $\epsilon = -0.125$, there will be a final further second-order transition to an unmagnetized state.

We conclude the presentation of this case with the coordinates of the relevant points for $K_2 = - 0.14$. They are:
\begin{itemize}
  \item canonical critical (CC) point: $K_{1} \backsimeq -0.3590, T  \backsimeq 0.3470, \epsilon  \backsimeq -0.1004$.
  \item canonical critical end (CCE) point: $K_{1} \backsimeq -0.4087, T  \backsimeq 0.2594, \epsilon  \backsimeq -0.08833$.
  \item microcanonical critical (MC) points: $K_{1} \backsimeq -0.4052, T  \backsimeq 0.2188, \epsilon  \backsimeq -0.1060$; $K_{1} \backsimeq -0.4577, T  \backsimeq 0.0834, \epsilon  \backsimeq -0.1404$.
  \item microcanonical critical end (MCE)  point: $\!\! K_{1} \backsimeq -0.4565, T  \backsimeq 0.08786, \epsilon  \backsimeq -0.1409$. 
   \item microcanonical triple (M3) point: $K_{1} \backsimeq -0.4561, T  \backsimeq 0.07516; 0.08618; 0.2294, \epsilon  \backsimeq  -0.1412$.  
 \end{itemize}

\begin{figure}[htbp]
\begin{center}
\begin{tabular}{cc}
 \includegraphics[clip, trim=3.7cm 8.5cm 3.7cm 8.5cm, width=0.5\textwidth]{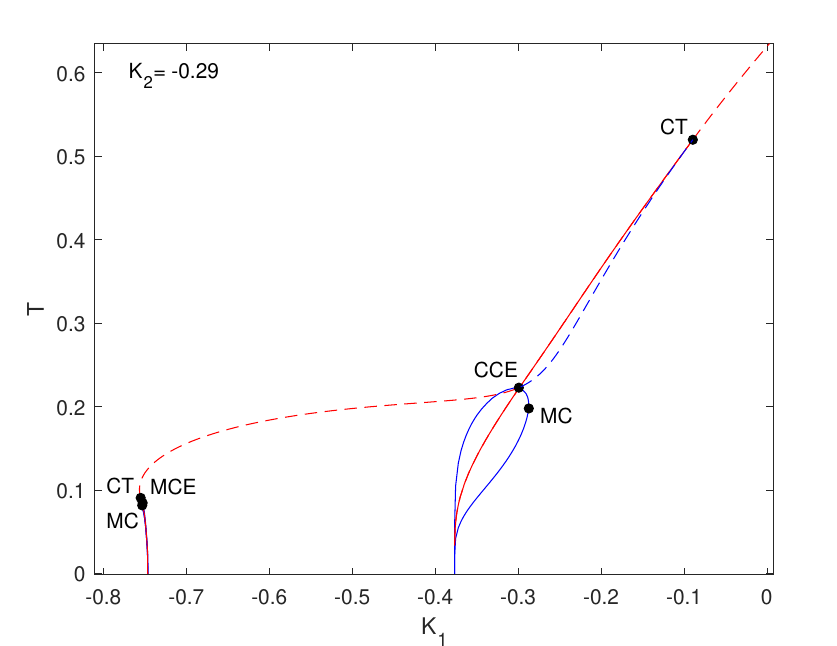} &
\includegraphics[clip, trim=3.7cm 8.5cm 3.7cm 8.5cm, width=0.5\textwidth]{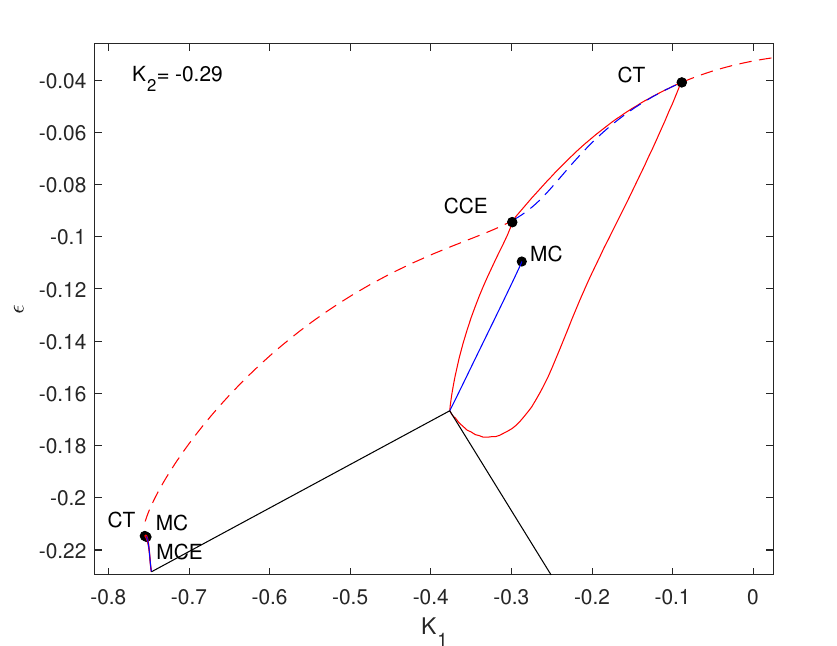}
\end{tabular}
\caption {The $(K_1,T)$ (left) and $(K_1,\epsilon)$ (right) phase diagrams of the canonical and microcanonical ensembles at $K_2=-0.29$. In the canonical case both first-order lines starting at $T=0$ end at a critical point CT, and a second-order line goes from one tricritical point CT to the critical end point CCE. In the microcanonical case both first-order lines starting at the ground energy end at a critical point MC, and there is only a second-order line, starting at a critical end point MCE and extending indefinitely. The small structure with the very close points CT, MC and MCE is detailed in the next figures with zoomed plots.}
\label{fig6}
\end{center}
\end{figure}

\begin{figure}[htbp]
\begin{center}
\begin{tabular}{cc}
 \includegraphics[clip, trim=3.7cm 8.5cm 3.7cm 8.5cm, width=0.5\textwidth]{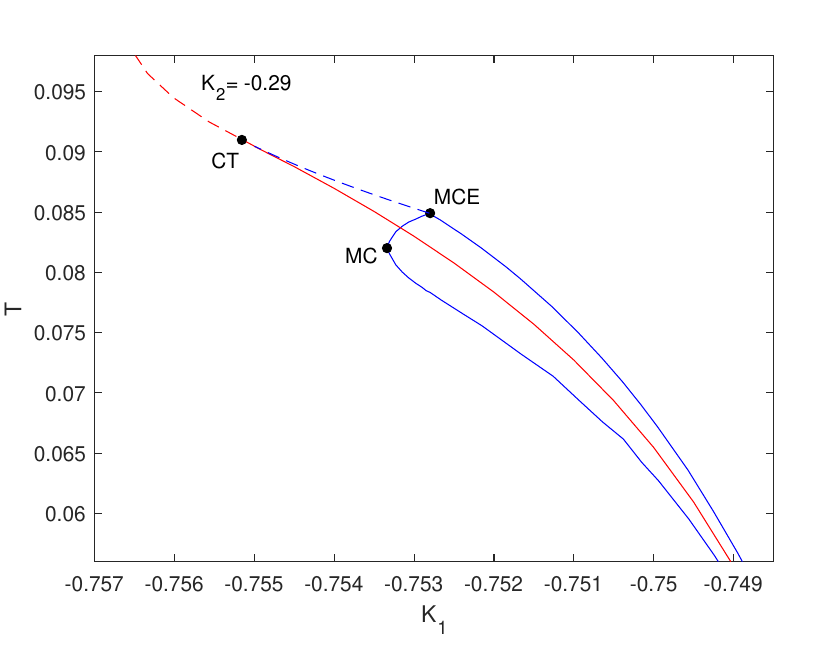} &
 \includegraphics[clip, trim=3.7cm 8.5cm 3.7cm 8.5cm, width=0.5\textwidth]{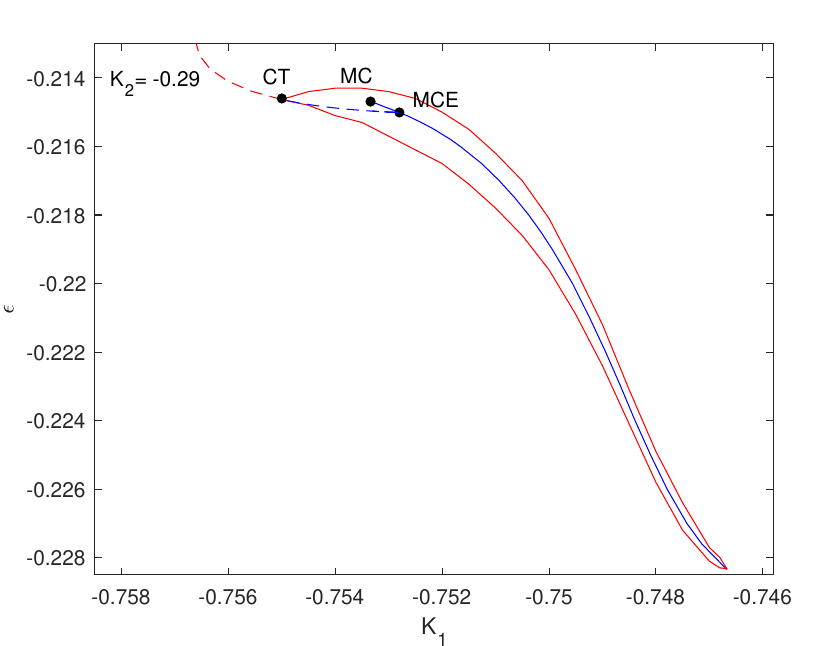}
\end{tabular}
\caption {The zoomed plots of the small region of the microcanonical $(K_1,T)$ (left) and $(K_1, \epsilon)$ (right) phase diagrams, at $K_2=-0.29$, around the MCE. The small structure, with a MC point and a MCE point very close, is similar to the one found for $K_2=-0.14$. It persists also at smaller values of $K_2$.}
\label{fig7}
\end{center}
\end{figure}

We pass now to the phase diagrams for $K_2 = - 0.29$. Also in this case we present in a figure the full $(K_1,T)$ and
$(K_1,\epsilon)$ phase diagrams, and in another figure a zoomed region of the two diagrams. The situation occurring for the canonical case in the $(K_1,T)$ diagram was already presented in Ref. \cite{jpa2019}: from figure \ref{fig6} (left) we see that there are two first-order lines starting at $T=0$. Both lines end at a tricritical point CT, at which a second-order line begins. One of these critical lines continues indefinitely, while the other ends at a critical end point CCE, where it meets the other first-order line.
The section of the right first-order line between $T=0$ and the CCE point separates two magnetized states. In the $(K_1,\epsilon)$ diagram, in figure \ref{fig6} (right), where the canonical first-order lines have two branches, those are clearly visible in the full diagram only for one of the two first-order lines, the one at the larger values of $K_1$. We can see the other in the zoomed plot in the right panel of  
figure \ref{fig7}.

In the microcanonical case we have still another kind of structure. We see in figure \ref{fig6} (right) that the first-order line starting at the ground energy at a $K_1$ value somewhat larger than $-0.4$, and separating two magnetized states, ends with a critical point, at variance with the canonical first-order line, that as seen in the left panel ends at a tricritical point. As a consequence, the microcanonical second-order line arriving from large values of $K_1$ and $\epsilon$ arrives up to the other short first-order line, starting at the ground energy at a $K_1$ value between $-0.8$ and $-0.7$. This last line is not visible in figure \ref{fig6}, but only in the zoomed plots in figure \ref{fig7}. We see in the panels of this figure that also this line ends with a critical point MC, and that the second-order line meets it at a critical end point MCE. This small structure, with a MC point and a MCE point very close, is similar to the one found in the previous case ($K_2=-0.14$), and thus it has the peculiarity to persist for quite a large range of $K_2$; in fact, we have found that it is present also at smaller values of $K_2$.

The coordinates of the relevant points of the phase diagrams for $K_2 = -0.29$ are: 
\begin{itemize}
  \item canonical tricritical (CT) points: $K_{1} \backsimeq -0.0890, T \backsimeq 0.5208, \epsilon \backsimeq -0.04090$ and $K_{1} \backsimeq -0.7550, T \backsimeq 0.0905, \epsilon  \backsimeq -0.2146$. 
  \item canonical critical end (CCE) point: $K_{1} \backsimeq -0.2990, T \backsimeq 0.2225, \epsilon \backsimeq -0.09450$.
  \item microcanonical critical (MC) points: $K_{1} \backsimeq -0.7533, T \backsimeq 0.08202, \epsilon \backsimeq -0.2147$ and  $K_{1} \backsimeq -0.2873,
	T \backsimeq 0.1981, \epsilon  \backsimeq -0.1095$.
  \item microcanonical critical end (MCE)  point: $\!\! K_{1} \backsimeq -0.7528, T \backsimeq 0.08492, \epsilon \backsimeq -0.2150$. 
 \end{itemize}

\begin{figure}[htbp]
\begin{center}
\begin{tabular}{cc}
 \includegraphics[clip, trim=3.5cm 8.5cm 3.7cm 8.5cm, width=0.49\textwidth]{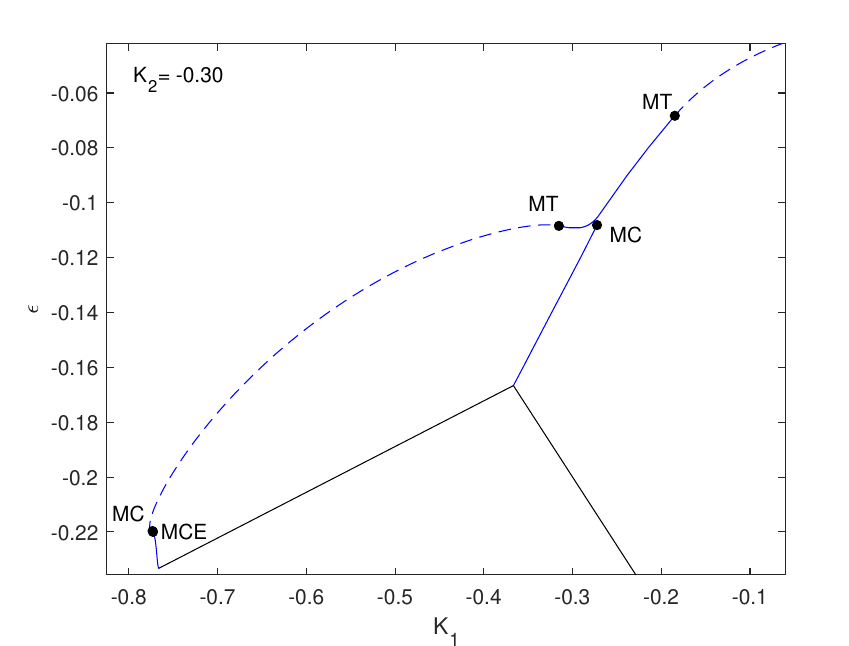} &
 \includegraphics[clip, trim=3.5cm 8.5cm 3.7cm 8.5cm, width=0.49\textwidth]{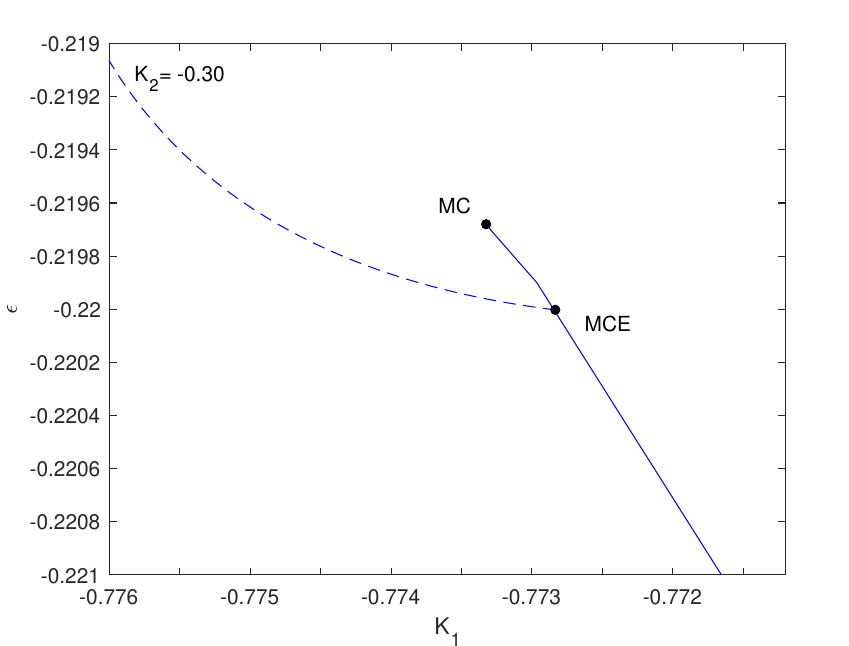}
\end{tabular}
\caption {The $(K_1, \epsilon)$ phase diagram of the  microcanonical ensemble at $K_2=-0.30$. The left plot shows the full diagram, while the right plot is the zoom of the region with the close points MC and MCE.}
\label{fig8}
\end{center}
\end{figure}

The final case we present is the one for $K_2 = -0.30$, for which we focus only on the microcanonical phase diagram. In figure \ref{fig8} we show, on the left, the full phase diagram, and, on the right, a zoomed region where a critical point MC and a critical end point MCE are very close. The difference with respect to the microcanonical diagram of the former case ($K_2=-0.29$) is given by the fact that the second-order line starting at the MCE (see in particular the right plot of the zoomed region) has a brief first-order line, between two tricritical points MT, included in it. The critical point MC, ending the first-order line separating two magnetized states, is very close to this first-order section. The further peculiarity of this diagram is that the first-order line between the two tricritical points is not monotonic seen as $\epsilon$ as a function of $K_1$, although this is not very evident at the scale of the full diagram. Furthermore, the $\epsilon$ coordinate of the MC point close to the first-order section within the second-order one is higher then the $\epsilon$ coordinate of the first tricritical point MT (see below the list of the coordinates of the relevant points). As a consequence, increasing $K_1$, starting e.g. at $K_1 = -0.6$ for a fixed $\epsilon$ value smaller than that of the MT but higher than that of the minimum of the first-order section, we pass with a continuous transition from an unmagnetized state to a magnetized state, then with a first-order transition again to the unmagnetized state, then with a first-order transition again to the magnetized state, and finally with a first-order transition to another magnetized state. 

The coordinates of the relevant points of the phase diagram for $K_2=-0.30$ are: 
\begin{itemize}
  \item microcanonical tricritical (MT) points: $K_{1} \backsimeq -0.3154, \epsilon  \backsimeq -0.1084$ and  $K_{1} \backsimeq -0.1848, \epsilon  \backsimeq -0.06830$.
  \item microcanonical critical (MC) points: $K_{1} \backsimeq -0.7733, \epsilon  \backsimeq -0.2197$ and  $K_{1} \backsimeq -0.2724,  \epsilon  \backsimeq -0.1082$.
  \item microcanonical critical end (MCE)  point: $K_{1} \backsimeq -0.7728, \epsilon \backsimeq -0.2200$.   
 \end{itemize}

\section{Normal form for a fourth-order critical point}
\label{normalform}
In this Section we present the normal form for a fourth-order critical point, that manifests itself, as we will explain, in our results. Our aim is to give a relatively
simple description, referring the reader to the cited literature for more detailed and more rigorous presentations.

\subsection{Codimension of singularities}
Full details of what briefly described in this Section, concerning the relation between different kinds of singularities in the thermodynamic functions and the different
relevant points, can be found in Ref. \cite{bb2005}.

The full thermodynamic phase diagram of our system is three-dimensional. In fact, the equilibrium state depends on the value of the coefficients $K_1$ and $K_2$ in the
Hamiltonian, and on the control variable identified by the energy per particle $\epsilon$ in the microcanonical ensemble and by the temperature $T$ in the canonical
ensemble. As shown in Ref. \cite{bb2005}, for a non-additive system, by studying the microcanonical entropy it is possible to determine the location, in the phase diagram,
of the phase transitions, both in the microcanonical ensemble itself and in the canonical one; due to inequivalence, the locations will in general be different in both ensembles.
The microcanonical entropy depends on the control variable, i.e., the energy per particle $\epsilon$,
and on the parameters of the Hamiltonian, $K_1$ and $K_2$ in our case, so that we can write $s=s(\epsilon,K_1,K_2)$. The phase transitions are associated to the singularities
of this function, with different types of singularities corresponding to different kinds of phase transitions in the two ensembles. The singularities occur when given
conditions are realized in $s=s(\epsilon,K_1,K_2)$ and/or in its derivatives with respect to $\epsilon$ (see Ref. \cite{bb2005} for details). It is not difficult
to argue that, loosely speaking, the number of conditions determines the dimension of the hypersurface, in the 3D phase space, associated to the singularity.
Therefore, we can have singularities associated to hypersurfaces of dimension $2$, $1$ and $0$ (the latter being obviously single points). Adopting the terminology
employed in \cite{bb2005}, a singularity is said to be of codimension $n$ if the associated hypersurface is of dimension $2 - n$, the number $2$ being the number of parameters
in the Hamiltonian. The singularities mentioned above are therefore identified as having codimension $0$, $1$ and $2$ respectively.

In this perspective, one can understand that the change of structure observed in different 2D cuts of the phase diagram, with the cuts determined, in our case, by the value
of $K_2$, is a consequence of the different regions that are included in the cut for different values of $K_2$. Other things that can be easily understood are the following.
First, the intersection of two codimension $0$ singularities (i.e., of two 2D surfaces) will occur in the 1D region associated to a codimension $1$ singularity, while the intersection
of two codimension $1$ singularities (or, equivalently, of three codimension $0$ singularities) will be at the single point determining a codimension $2$ singularity.
Secondly, one can expect to have, in a 2D cut of the phase diagram, the presence of codimension $0$ singularities, in this case represented by 1D lines (the intersection
of the 2D surface with the 2D cut), and of codimension $1$ singularities, represented by single points in the 2D cut. Generically, however, we will not have
a codimension $2$ singularity in a 2D cut, since this singularity is associated to a precise point in the 3D diagram.

In non-additive systems characterized, in their equilibrium states, by the value of an order parameter, it is generally the case that the thermodynamic function associated to the
ensemble under study is obtained by an extremization problem. In our case, where we have used the Hubbard-Stratonovich transformation and the min-max procedure,
equations (\ref{freehs}) and (\ref{minmaxmic}) give, respectively, the canonical free energy and the microcanonical entropy. Let us adopt the symbol $\phi(\beta)$ for the function
$\beta f(\beta)$; rewriting Eq. (\ref{freehs}) with the explicit indication of the dependence on the Hamiltonian parameters $K_1$ and $K_2$, we have
\be
\label{freehsb}
\phi(\beta,K_1,K_2) = \min_x [\widetilde{\phi}(\beta,x,K_1,K_2)] \, .
\ee
Analogously, we rewrite Eq. (\ref{minmaxmic}) as
\bea
\label{minmaxmicb}
s(\epsilon,K_1,K_2) &=& \max_x \left\{ \min_\beta \left[ \beta \epsilon - \widetilde{\phi}(\beta,x,K_1,K_2) \right] \right\} \nonumber \\
&\equiv& \max_x \left[\widehat{s}(\epsilon,x,K_1,K_2)\right] \, ,
\eea
the second equality defining the function $\widehat{s}(\epsilon,x,K_1,K_2)$.
In these optimization problems, the auxiliary variable $x$ takes in principle the values from $-\infty$ to $+\infty$, and the various values of $x$ do not have a direct physical interpretation. However, it is possible to show \cite{Campa2014} that the value of $x$ that satisfies the optimization problems is equal to the equilibrium value of $m$, whose absolute value is between $0$ and $1$. It is important to stress the following. The function $\tilde{\phi}(\beta,x,K_1,K_2)$ is analytic, and the possible singularities of the function
$\phi(\beta,K_1,K_2)$ in Eq. (\ref{freehsb}) arise from the minimization procedure with respect to $x$. In a similar way the possible singularities of the function
$s(\epsilon,K_1,K_2)$ in Eq. (\ref{minmaxmicb}) arise from the maximization with respect to $x$; in fact, the minimization with respect to $\beta$ is a Legendre-Fenchel
transform of the function $\tilde{\phi}(\beta,x,K_1,K_2)$, which, besides being analytic, is also convex with respect to $\beta$ (being minus the logarithm of a partition
function \cite{jpa2019}), so that the function
$\widehat{s}(\epsilon,x,K_1,K_2)$ in the last member of Eq. (\ref{minmaxmicb}) is analytic.

Let us compare the last expression with the one that we would have in the case in which the microcanonical entropy is obtained by a direct counting procedure \cite{jsp2021}:
\begin{equation}
\label{maxtilds}
s(\epsilon,K_1,K_2) = \max_m \left[\widetilde{s}(m,\epsilon,K_1,K_2)\right] \, .
\end{equation}
In this expression the function $\widetilde{s}(m,\epsilon,K_1,K_2)$ is the logarithm of the number of states with given energy $\epsilon$ and given magnetization $m$.
Apart from very simple models, already in a system like ours the computation of $\widetilde{s}(m,\epsilon,K_1,K_2)$ by direct counting is rather cumbersome, so that
the min-max procedure is considerably preferable \cite{jsp2021}. Also in this case the function $\tilde{s}(m,\epsilon,K_1,K_2)$ is analytic, and the singularities stem from
the maximization procedure.
The point that interests us here is the following. In the case of direct counting, the variable $m$ over which one has to
optimize is the magnetization for any value between $-1$ and $1$, therefore also for values different from that satisfying the optimization problem (i.e., for non-equilibrium
values of the magnetization); this is at variance
with the extremization problem with respect to the auxiliary variable $x$. However, the very fact that, as underlined above, the value of $x$ that realizes the extremum, let us call
it $x^*$, is equal to the equilibrium value $m^*$ of magnetization $m$, makes it possible to study the function $\widehat{s}(\epsilon,x,K_1,K_2)$ for $x$ in the small neighborhood of $x^*$ assuming that in this neighborhood $x$ represents the magnetization $m$. In other words, the study of $\widehat{s}(\epsilon,x,K_1,K_2)$ around $x=x^*$ gives the same information of the study of $\widetilde{s}(m,\epsilon,K_1,K_2)$ around $m=m^*$. In particular, this is true for the study of the second-order transitions that occur at $m^*=0$.
Therefore, in the formal procedure that will be described in this section, in which we do not need the explicit expression of the functions, for definiteness we will consider a function $\widetilde{s}(m,\epsilon,K_1,K_2)$.

We also note that the whole analysis that will be carried out in the rest of this section can equally be done for the canonical case, with the function
$\widetilde{\phi}(\beta,x,K_1,K_2)$ replacing the function $\widetilde{s}(m,\epsilon, K_1, K_2)$, and the minimization replacing the maximization. Obviously, with the
function $\widetilde{\phi}(\beta,x,K_1,K_2)$ one would study the singularities in the canonical phase diagram.

\subsection{The fourth-order critical point}
In this Section we explain where the fourth-order critical point is located in our phase diagram and we obtain its normal form description. We stress that our procedure
is intentionally simple, and that the reader can found more rigorous treatments in the literature. Concerning in particular critical end points, we refer to some
older \cite{ishi1991,helena1994} and some more recent works \cite{sumedha2020,mukher2023}.
Here our main purpose is to show in a simple way that, while in general it is not possible to use a Landau expansion for the description of a critical end point,
this becomes possible near a fourth-order critical point, where the magnetization at the critical end point is vanishingly small.

We are interested to a particular singularity at $m^*=0$, a fourth-order critical point. This is a codimension $2$ singularity, that occurs when, in the 3D phase diagram, three 1D lines
associated to three codimension $1$ singularities meet at a point. These three lines are: a line of tricritical points, a line of critical points, and a line of critical end points.
We have not explicitly shown a fourth-order point in one of our graphs. As we have explained above, this would require to determine the exact value of $K_2$ at which to plot
the 2D cut with coordinates $(K_1,\epsilon)$. However, the comparison of the right plots of figure 1 and figure 2 indicates that this value of $K_2$ is between $-0.10$ and $0$.
In fact, in figure 1 we have a tricritical point, while in figure 2 we see a critical point and a critical end point. A similar passage was observed in Ref. \cite{pre2019}
for the BEG model with a negative coupling coefficient for the biquadratic interaction. Since we have to study the function $\widetilde{s}(m,\epsilon,K_1,K_2)$ in the neighborhood
of $m=0$, and knowing that this function is analytic, we expand it in powers of $m$. The function is even in $m$, since it reflects the symmetry $m \to -m$ in the Hamiltonian
of the model, and thus there are only even powers of $m$ in the expansion. Let us then denote
\begin{equation}
\label{defder}
h_j (\epsilon,K_1,K_2) = \frac{1}{(2j)!}\left.\frac{\partial^{2j} \widetilde{s}}{\partial m^{2j}}\right|_{m=0} \, .
\end{equation}
For small $m$ we write
\begin{equation}
\label{powerexp}
\widetilde{s}(m) = h_0 + h_1 m^2 + h_2 m^4 + h_3 m^6 + h_4 m^8 + O(m^{10}) \, .
\end{equation}
Here and in the following for brevity we do not write explicitly the dependence on the energy per particle $\epsilon$ and on the Hamiltonian parameters $K_1$ and $K_2$.
As will be shown shortly, from this expression it is possible to obtain the normal form of $\widetilde{s}$ that describes the fourth-order critical point.
In particular, we will see why this singularity requires an expansion up the to eighth power of $m$.

The expansion (\ref{powerexp}) can describe a phase transition only if this transition involves the values of $\widetilde{s}$ for very small $m$.
In particular, what count are the maxima, in $m$, of $\widetilde{s}$. Thus, if the expansion (\ref{powerexp}) has a relevance in the study of a singularity of
$s(\epsilon,K_1,K_2)$, it is necessary that in the range of its validity, i.e., for small values of $m$, there is a maximum.
Obviously, the presence of such a maximum does not exclude that, outside the range of validity of (\ref{powerexp}), i.e., for larger values of
$m$, the function $\widetilde{s}$ has a higher maximum, so that the one at small $m$ is only local.
This observation allows to see which types of phase transition can be described and represented, generically, with (\ref{powerexp}). This can be done conveniently
by considering singularities of increasing codimension. Beginning with codimension $0$ singularities, we note that, generically, systems with symmetry in the
Hamiltonian, like ours, have a continuous phase transition at $m=0$ associated to symmetry breaking, and this
is the only singularity of codimension $0$ involving only states with $m=0$ or very small $m$ \cite{bb2005} (a first-order phase transition between a state with $m=0$
and a state with positive $m$ is of codimension $0$, but besides the state with $m=0$ it involves another state with, generically, $m$ positive and not small). This singularity
can be described by a normal form like $\widetilde{s}(m) = -m^4 + a m^2$,
with $a$ small in absolute value. In fact, at the symmetry breaking phase transition the order parameter goes continuously from $m=0$ to $m>0$ (we can consider just $m\ge 0$,
because of the symmetry of the Hamiltonian). The mentioned normal form has a maximum at $m=0$ for $a<0$, and at positive and small $m$ for $a$ positive and small.
This normal form can be obtained by the expansion (\ref{powerexp}) by neglecting the term $h_0$, irrelevant for the behavior with respect to the variation of $m$, and keeping only
the terms proportional to $m^2$ and $m^4$. The latter has to be negative to have a maximum at $m=0$ or at a small $m$ value. Redefining the coefficient we arrive at the
above normal form. The phase transition occurs then at $a=0$. Since this coefficient depends on $(\epsilon,K_1,K_2)$, we see that, in agreement with what described at the
beginning of this section, this singularity defines a 2D surface in the three-dimensional phase diagram.

Going to codimension $1$ singularities, it is possible to see that, generically, only a tricritical point involves only states with $m=0$ or very small $m$. The other
codimension $1$ singularities, like critical points, triple points and critical end points, involve, generically, also states with $m$ positive and not small. The normal form
associated to a tricritical point can be written as \cite{bb2005}
\begin{equation}
\label{normtricr}
\widetilde{s}(m) = -m^6 -\frac{3}{2}bm^4 -3 am^2 \, ,
\end{equation}
with $a$ and $b$ small in absolute value. Comparing with (\ref{powerexp}) we see that, again neglecting the term $h_0$, it is obtained by redefining $h_1$, $h_2$ and
$h_3$, setting the latter equal to $-1$. The coefficient of $m^6$ has to be negative in order to have the absolute maximum of $\widetilde{s}(m)$ at $m=0$ or at a small
value of $m$. The singularity occurs at $a=b=0$ (or equivalently $h_1=h_2=0$); these two equalities define a 1D line in the phase diagram, in agreement with what explained
above for codimension $1$ singularities.

Finally we consider codimension $2$ singularities, the purpose of this section. From the above analysis, it is natural to guess that
the condition $h_1=h_2=h_3=0$, with $h_4<0$, determines such a singularity. However, not all of them will be described in this way, as we argue
shortly. Before, we note that the condition $h_1=h_2=h_3=0$ determines isolated points in the 3D phase diagram, and therefore in a 2D cut of the diagram,
obtained e.g., by fixing the value of $K_2$, generically these points are not shown, since they could occur only at precise values of $K_2$. It remains to see which kind
of codimension $2$ singularities can be described by the vanishing of $h_1$, $h_2$ and $h_3$, and which are the associated kinds of structure changes in the 2D cuts defined,
e.g., by a fixed value of $K_2$.

As noted above, the change of structure of the 2D phase diagram is characterized by the meeting or departing of points (the
intersection of 1D lines with the 2D plot) representing codimension $1$ singularities (or when such a point reaches the ground energy or the
temperature $T=0$). Therefore, we have to consider the meeting or departing of such singularities. Preliminary, we have to reconcile this
observation with the above conclusion that, as far as codimension $1$ singularities are concerned, the normal form (\ref{powerexp}) can describe,
generically, only a tricritical point. So, how can we describe the meeting of codimension $1$ singularities some of which are not tricritical
points? The reason resides in the word ``generically''. As we noted, critical end points and critical points are associated, generically, to transitions
that involve states with a not small magnetization $m$. However, it may happen that the finite magnetization of such states becomes small, and this is
precisely what happens when a critical end point or a critical point get very close to a tricritical point. Therefore, we conclude that in the
3D phase diagram the point determined by $h_1=h_2=h_3=0$, representing the codimension $2$ singularity, is the meeting point of three different lines
each one of them associated to a codimension $1$ singularity: a line of tricritical points, a line of critical points, and a line of critical end
points. This particular point can be termed fourth-order critical point \cite{pre2019}. On the other hand, we have to exclude a possible line of triple
points. In fact, a triple point involves two states with finite magnetization, and a description with a form like (\ref{powerexp}) would require
that both of these finite magnetizations approach zero, and we expect this to be associated to singularities of higher order. Therefore a codimension $2$
singularity involving a triple point will not be described by an expression like (\ref{powerexp}).

In conclusion, the condition $h_1=h_2=h_3=0$ with $h_4<0$ represents the codimension $2$ singularity given by the meeting of the three lines just
mentioned, and the neighborhood of the singularity will be described by small values of these three coefficients. The structure of a 2D cut of the
phase diagram defined by a given value of $K_2$ will depend on the orientation of this plane with respect to the three lines.
We come back to this point below.

Since Eq. (\ref{powerexp}) has to be studied for small values of $h_1$, $h_2$ and $h_3$, and for $h_4<0$, we can consider fixed the value of $h_4$ and we can
normalize the expression dividing it by the absolute value of $h_4$. By further neglecting the irrelevant term $h_0$, we arrive, redefining
the other coefficients, at the normal form
\begin{equation}
\label{normsec}
\widetilde{s}(m) = -m^8 + h_3 m^6 + h_2 m^4 + h_1 m^2 \, .
\end{equation}
We recall that the coefficients $h_1$, $h_2$ and $h_3$ are functions of $(\epsilon,K_1,K_2$), and we emphasize again that this expression represents
the function $\widetilde{s}$ in the vicinity of the codimension $2$ singularity only when $m$, $h_1$, $h_2$ and $h_3$ are small in absolute value.

It is convenient to make a simple rescaling of the coefficients $h_1$, $h_2$ and $h_3$, defined by: $h_1 = -4a$, $h_2 = -2b$, $h_3 = 4c/3$; so the normal
form can be written as
\begin{equation}
\label{normsec1}
\widetilde{s}(m) = -m^8 +\frac{4}{3}cm^6 -2bm^4 -4am^2 \, .
\end{equation}
Since the coefficients $a$, $b$ and $c$ are functions of $(\epsilon,K_1,K_2)$ and they vanish at the codimension $2$ singularity, then, taking
the origin of coordinates for $(\epsilon,K_1,K_2)$ exactly at the singularity, in its neighborhood one can make the linear approximation
\begin{eqnarray}
\label{powera}
a &=& \gamma_{11}\epsilon + \gamma_{12}K_1 + \gamma_{13}K_2 \\
\label{powerb}
b &=& \gamma_{21}\epsilon + \gamma_{22}K_1 + \gamma_{23}K_2 \\
\label{powerc}
c &=& \gamma_{31}\epsilon + \gamma_{32}K_1 + \gamma_{33}K_2
\end{eqnarray}

We find convenient to study the behavior of the normal form by considering 2D planes in which the coefficient $c$ is fixed (within the linear approximation
defined by (\ref{powera}-\ref{powerc})). To do this, we proceed as follows.
Defining the vectors $\boldsymbol{\gamma}_i = (\gamma_{i1},\gamma_{i2},\gamma_{i3}), \,\,\,\,\, i=1,2,3$, we have that
the vectors $\boldsymbol{\gamma}_1$, $\boldsymbol{\gamma}_2$ and $\boldsymbol{\gamma}_3$ are, respectively, parallel to the normal to the planes defined
by $a=0$, $b=0$ and $c=0$. These vectors are generally independent and not mutually perpendicular. Then, we perform an orthogonal transformation,
in the 3D phase diagram, defined by an orthogonal matrix ${\rm O} \equiv \{O_{ij}\}$:
\begin{eqnarray}
\label{orthoa}
\epsilon &=& O_{11}x + O_{12}y + O_{13}z \\
\label{orthob}
K_1 &=& O_{21}x + O_{22}y + O_{23}z \\
\label{orthoc}
K_2 &=& O_{31}x + O_{32}y + O_{33}z \, .
\end{eqnarray}
Taking one of the three orthogonal unit vectors, defining the orthogonal transformation, parallel to the vector $\boldsymbol{\gamma}_3$,
the relations (\ref{powera}-\ref{powerc}) expressed in function of $(x,y,z)$ become
\begin{eqnarray}
\label{powerap}
a &=& \gamma_{11}'x + \gamma_{12}'y + \gamma_{13}'z \\
\label{powerbp}
b &=& \gamma_{21}'x + \gamma_{22}'y + \gamma_{23}'z \\
\label{powercp}
c &=&  \gamma_{33}'z
\end{eqnarray}
Since in general the vectors $\boldsymbol{\gamma}_i$s are not mutually perpendicular, the matrix elements $\gamma_{13}'$ and $\gamma_{23}'$ will
in general be different from zero.

Let us now take a plane, in our 3D phase diagram, defined by a given (small) value of $z$. By virtue of the last relations, in this plane the value
of $c$ will be constant (we recall that we are always reasoning in a small neighborhood of the singularity), and equal to $\gamma_{33}'z$.
Therefore, restricting the analysis to this plane, in the study of the normal form (\ref{normsec1}) it is possible to consider fixed the value
of $c$. Denoting this fixed value with $c_0$, we then have to study the normal form
\begin{equation}
\label{normsec2}
\widetilde{s}(m) = -m^8 +\frac{4}{3}c_0m^6 -2bm^4 -4am^2
\end{equation}
for a given fixed value of $c_0$ (small in absolute value), in a neighborhood of $(a,b)=(0,0)$. For the interested reader
few details of the calculations are shown in Appendices A and B; here we present the results, considering separately
the cases $c_0>0$ and $c_0<0$.

\subsection{The $(a,b)$ phase diagram for $c_0>0$}

We start with the codimension $1$ singularities, that in the plane are single points.

\begin{itemize}
\item{
There is a critical end point at $(a,b)=(0,\frac{2}{9}c_0^2)$.
}
\item{
There is a critical point at $(a,b)=(-\frac{1}{27}c_0^3,\frac{1}{3}c_0^2)$.
}
\end{itemize}

Then, we consider the codimension $0$ singularities, that in the plane are lines.

\begin{itemize}
\item{
There is a line of second-order transitions from an unmagnetized $m=0$ state to a magnetized state; the line is defined in
the $(a,b)$ plane by: $a=0$, $b\ge \frac{2}{9}c_0^2$.
}
\item{
There is a line of first-order transitions that is met from the line of second-order transitions at the critical end point.
The line is defined as follows. For $\frac{2}{9}c_0^2 \le b \le \frac{c_0^2}{3}$ it is given by $a=\frac{c_0}{27}(2c_0^2-9b)$, while for
$b \le \frac{2}{9}c_0^2$ it is given by
$a=\frac{1}{729}\left[32c_0^3-162bc_0 +\frac{1}{2}\left(16c_0^2-54b\right)^{\frac{3}{2}}\right]$. The latter portion of
the line marks a first-order transition between an unmagnetized state and a magnetized state, while the former
portion corresponds to a first-order transition between two different magnetized states. The two portions join at
the critical end point, where they have the same slope, with $\frac{\partial a}{\partial b} =-\frac{c_0}{3}$.
This first-order line ends at the critical point, in the portion corresponding to the transition between two different
magnetized states.
}
\end{itemize}

\begin{figure}[!htp]
\centering
\includegraphics[scale=0.5,trim= 0cm 7cm 0cm 8cm,clip]{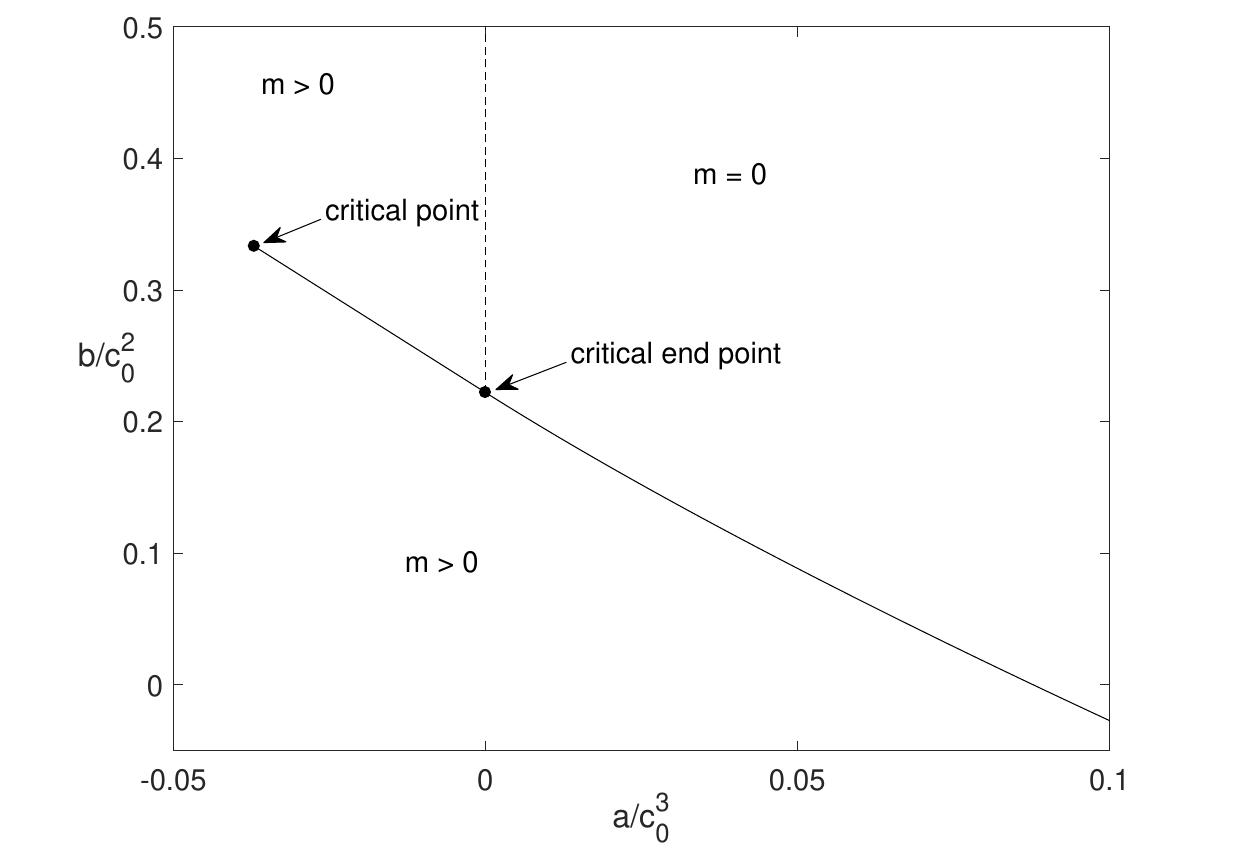}
\caption{Phase diagram in the $(a,b)$ plane, in the neighborhood of the origin, for $c_0>0$. The dashed line is a second-order
transition line, while the solid line is a first-order transition line, either between two magnetized states or between an
unmagnetized and a magnetized state. This line ends at a critical point on the side of the transition between two
magnetized states.}
\label{diagcpos}
\end{figure}

A plot of this $(a,b)$ phase diagram is given in figure \ref{diagcpos}.

\subsection{The $(a,b)$ phase diagram for $c_0<0$}

\begin{figure}[!htp]
\centering
\includegraphics[scale=0.5,trim= 0cm 7cm 0cm 8cm,clip]{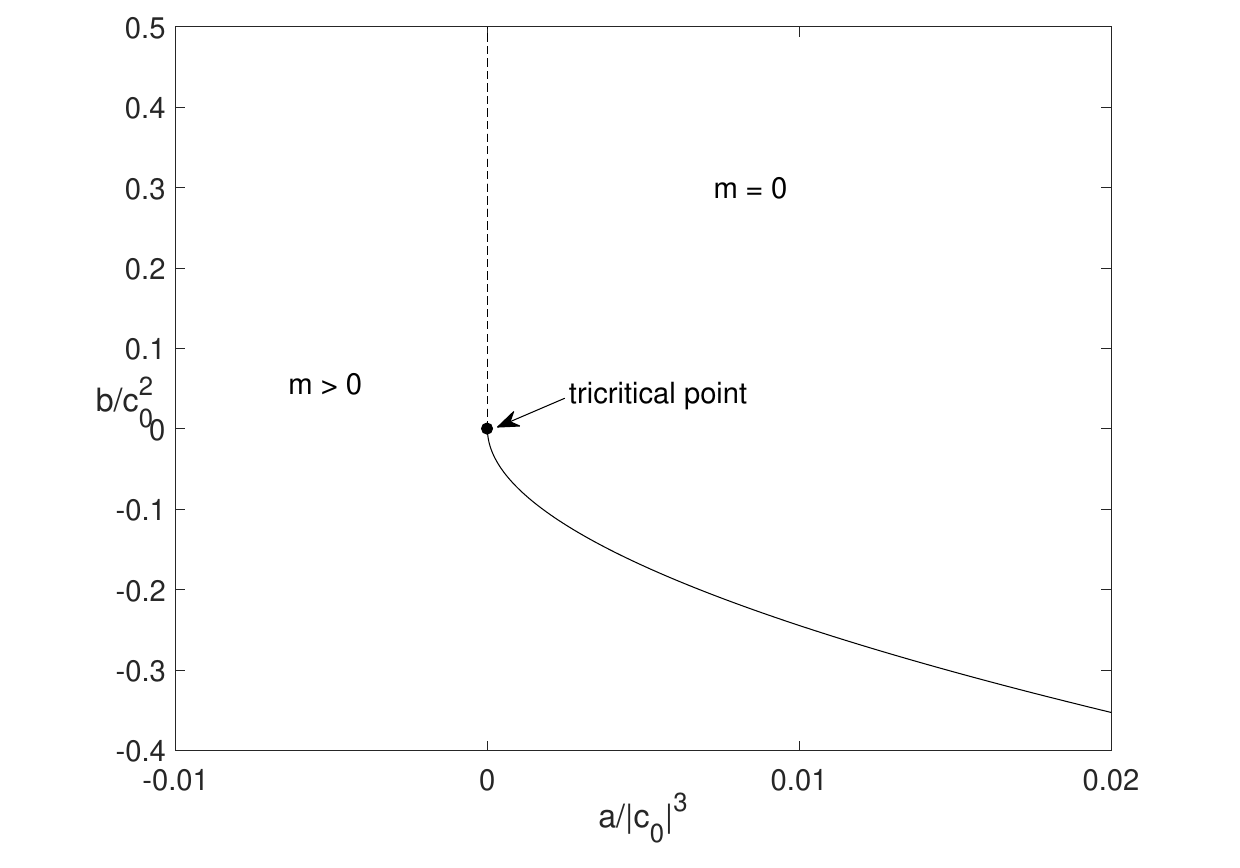}
\caption{Phase diagram in the $(a,b)$ plane, in the neighborhood of the origin, for $c_0<0$. The dashed line is a second-order
transition line, while the solid line is a first-order transition line between an
unmagnetized and a magnetized state. The two lines meet at the tricritical point at the origin.}
\label{diagcneg}
\end{figure}

There is only one codimension $1$ singularity:

\begin{itemize}
\item{
There is a tricritical point in the origin, $(a,b)=(0,0)$.
}
\end{itemize}
On the other hand, the codimension $0$ singularities are the following
\begin{itemize}
\item{
There is a line of second-order transitions from an unmagnetized $m=0$ state to a magnetized state; the line is defined in
the $(a,b)$ plane by: $a=0$, $b\ge 0$.
}
\item{
There is a line of first-order transitions between an unmagnetized state and a magnetized state, defined by
$a=\frac{1}{729}\left[32c_0^3-162bc_0 +\frac{1}{2}\left(16c_0^2-54b\right)^{\frac{3}{2}}\right]$, $b\le 0$. The value
of $a$ at $b=0$ is equal to zero, while for $b<0$ it is $a>0$. The line of first-order transitions meets the line of
second-order transitions at the tricritical point, where both lines have a vertical slope, $\frac{\partial a}{\partial b} =0$.
}
\end{itemize}
A plot of this $(a,b)$ phase diagram is given in figure \ref{diagcneg}.

It is not surprising that this structure is the same as that determined by the simpler normal form (\ref{normtricr}) considered
in Ref. \cite{bb2005}, that involves only up to $m^6$, with a coefficient $-1$, and the two parameters $a$ and $b$ (obviously
rescaled in a different way than what done here). In fact,
a negative coefficient of the term with $m^6$ makes not necessary another negative coefficient in the term with $m^8$. In any case,
in our study, since the term with $m^8$ is necessary when the term with $m^6$ has a positive coefficient, for coherence we
keep it also when the term with $m^6$ has a negative coefficient.

\subsection{The 2D phase diagrams near the codimension $2$ singularity}

Let us summarize what the results just mentioned imply for the structure of the phase diagram 
now go back to the relations (\ref{powera}-\ref{powerc}) between the coefficients $(a,b,c)$ and the phase diagram coordinates
$(\epsilon,k_1,k_2)$.

{\it In the neighborhood of the codimension $2$ singularity} the lines corresponding to the codimension $1$ singularities are determined by:
\begin{itemize}
\item{
The line of tricritical points by $a=0$, $b=0$, $c<0$;
}
\item{
The line of critical end points by $a=0$, $b=\frac{2}{9}c^2$, $c>0$;
}
\item{
The line of critical points by at $a=-\frac{1}{27}c^3$, $b=\frac{1}{3}c^2$, $c>0$.
}
\end{itemize}

We note the following things. First, looking at the relations defining the line of critical end points and that defining the line of critical
points, we see that they become tangent at the codimension $2$ singularity, where $c\to 0$. Second, in the coordinates $(a,b,c)$
the line of tricritical points is also tangent, at the codimension $2$ singularity, to the other two lines, although approaching the
singularity from the opposite side. However, third, as we have already remarked, the normal vectors to the planes
$a=0$, $b=0$, $c=0$, i.e., the vectors $\boldsymbol{\gamma}_i$, $\, i=1,2,3$, are not mutually orthogonal.

\begin{figure}[!htp]
\centering
\includegraphics[scale=0.6,trim= 0cm 9cm 0cm 8cm,clip]{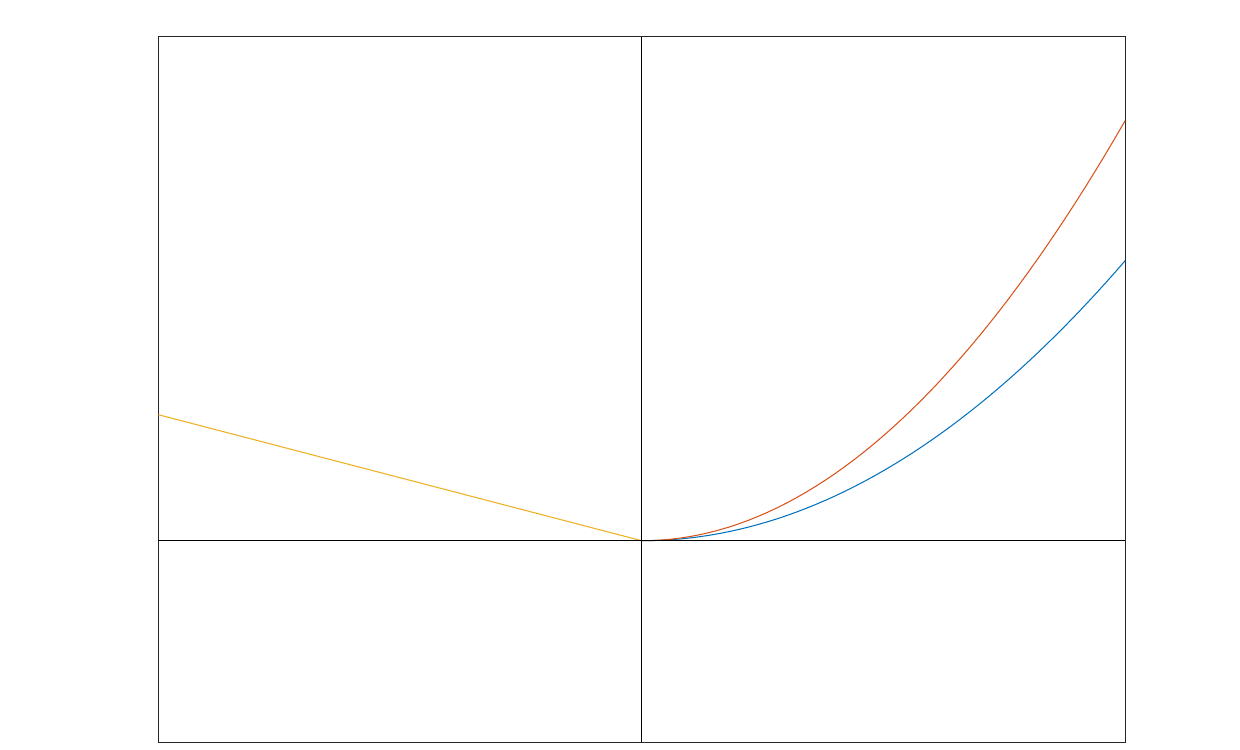}
\caption{Assumed structure of a generic 2D cut of the phase diagram in the vicinity of the codimension $2$ singularity, placed at the origin. The 2D cut is taken
to include exactly the two straight lines (the two on the right) to which two of the meeting lines tend when they approach the meeting point. The line on
the left is the projection on the 2D cut of the straight line to which the line of tricritical points tends while approaching the singularity.
The lines on the right are the projections of the line associated to the critical end points and of the line associated to the critical points.
}
\label{nodiag}
\end{figure}

Then one can envisage a plot like that in figure \ref{nodiag} for a generic 2D cut of the phase diagram described by the coordinates
$(\epsilon,K_1,K_2)$. In the plot, the three lines are projection on the given plane of the three lines associated to the codimension $1$
singularities. A member of a family of 2D cuts perpendicular to the plane depicted in the figure can include either a critical end point and a
critical point or a tricritical point. Other orientations for the family of 2D cuts can result in different cases for the appearance of
the codimension $1$ singularities in the members of the family.

\section{Conclusions}
\label{sec:conclusions}

In this paper we have studied the phase diagram of an Ising chain with competing interactions in the microcanonical ensemble. The model had been studied previously in the canonical ensemble~\cite{jpa2019}. We introduced the model in Section~\ref{sec:the_model}. Its interest relies in the presence of both a mean-field coupling $J$, a
nearest-neighbor coupling $K_1$ and a next-nearest-neighbor coupling $K_2$. The signs of the latter couplings can be chosen positive or negative, causing a competition with the ferromagnetic coupling $J$, which can be taken to be $J=1$ without loss of generality. The positive (ferromagnetic) mean-field coupling assures the presence of a phase transition for this one-dimensional system. The solution in the canonical ensemble had been obtained using the transfer matrix method and a saddle-point integration in order to perform the thermodynamic limit. Already in the canonical ensemble, the model was featuring eight different regions of parameters with qualitatively distinct phase diagrams in the $(K_1,T)$ plane at fixed values of $K_2$. 

Although the model is relatively simple, its microcanonical entropy is
hard to compute by using direct counting of the microstates. An alternative method had been proposed long ago \cite{Campa2009}
using the inversion of two limits: the minimum with respect to the inverse temperature and the maximum with respect to the Landau order parameter. This method was recently thoroughly discussed in Ref.~\cite{jsp2021}.
We have applied this method to our Ising chain revealing, with respect to the canonical solution, a different and even richer phase diagram and therefore confirming the presence of ensemble inequivalence. 

In Section \ref{microen} we have recapped the method to obtain the microcanonical entropy, and in Section \ref{microphase} we have performed a detailed analysis of the
$(K_1,T)$ and the $(K_1,\epsilon)$ phase diagrams for several values of $K_2$. The results have been presented and commented in details, and here we just would like to
underline the most peculiar feature that characterizes the phase diagrams. Concerning the diagrams obtained for $K_2=-0.14$, shown in their entirety in figure \ref{fig3} and
in a zoomed region in figure \ref{fig4}, we have seen in that a very small structure is present, so small that at the scale of figure \ref{fig3} it is not visible. In this structure, extending over very small ranges in $K_1$ and $\epsilon$, there are very close codimension $1$ singularities. In particular, in the right panel of figure \ref{fig3} we see the presence of a critical point (MC), a critical end point (MCE) and a triple point (M3) in the $(K_1,\epsilon)$ diagram. The MC point and the MCE point are very close, obviously, also in the $(K_1,T)$ diagram, in the left panel of the same figure. Moreover, for the M3 point, that in $(K_1,T)$ diagram is associated with three different temperatures, only two of those temperatures are close to the ones of the MC and the MCE points. The strangeness of this structure is also due to the following fact: while the microcanonical triple point M3 has not been found at the other $K_2$ values, nonetheless the presence of very close MC and MCE points has proved to be a robust feature. See for example figure \ref{fig7}, showing a zoomed small region of the phase diagram for $K_2=-0.29$, and the right panel of figure \ref{fig8}, referring to an analogous zoomed small region of the phase diagram for $K_2=-0.30$. We underline that this common feature in the microcanonical phase diagrams is accompanied by different structures in the canonical case: for $K_2=-0.14$ the small region is in the forbidden zone, while for $K_2=-0.29$ and $K_2=-0.30$ (although not explicitly shown in the latter case) the close MC and MCE points have replaced a canonical tricritical point.
Finally, we emphasize that we have found this small microcanonical structure also at smaller $K_2$ values (larger negative values), not shown here, justifying our statement that it appears to be very robust. 

We found that the regions of inequivalence typically occur at lower temperatures with respect to the case in which no short-range competing interactions are present. A tentative justification of this feature relies on the complexity of the ground states in presence of competing interactions. This complexity is partially preserved at finite temperature and disappears only at higher temperatures when the paramagnetic phase dominates in the phase diagram.      

The very simple structure of the phase diagram of the model with only nearest-neighbor interactions beside the mean-field one, i.e. $K_2=0$, is in striking contrast with the phase diagrams with $K_2<0$. Although the simpler $K_2=0$ model already presents ensemble inequivalence, as shown in figure \ref{fig1} (where we have rederived the results of Ref. \cite{Mukamel2005}), the diagram is characterized only by a tricritical point in both ensembles. This point appears for $K_1<0$, since the first-order phase transition occurs when the mean-field interaction and the nearest-neighbor interaction compete. The competition is still stronger in the model where the nearest-neighbor and the
next-nearest-neighbor interaction compete among them and each of them competes with the mean-field interaction, i.e., when $K_1$ and $K_2$ are both negative. A clear manifestation of this property is the presence of ferrimagnetic states, which are absent for $K_2=0$, in addition to ferromagnetic states. This occurs already in the ground states \cite{jpa2019}, and persists at positive temperatures.

Finally, in Section~\ref{normalform}, motivated by the presence of a fourth-order critical point in the phase diagram, we have presented an analysis of the general form of the entropy in the neighborhood of such points, extending the study performed in \cite{bb2005} to singularities of higher codimension. This has implied a careful study of the expansion of the entropy in terms of the powers of the order parameter up to order eight.

\section*{Acknolwledgements}
We thank N. Aninikian, N. Defenu and G. Gori for useful discussions. Support from the CNR/MESRA project
"Statistical Physics of Classical and Quantum Non Local Hamiltonians:
Phase Diagrams and Renormalization Group" is gratefully acknowledged. V.H. acknowledge the receipt of the  grant in the frame of the research  projects No. SCS 21AG-1C006 and No. SCS 23SC-CNR-1C006.

\appendix

\section{Details of the calculations for $c_0>0$}
\label{appcalcp}

In this Appendix we give few details of the calculations for the study of the normal form (\ref{normsec2}) in the $(a,b)$ plane,
where the value of $c_0$ is fixed and positive. For convenience in the following we remove the subscript of $c_0$, and we rewrite the normal form:
\begin{equation}
\label{entrabc}
\widetilde{s}(m) = -m^8 +\frac{4}{3}cm^6 -2bm^4 -4am^2 \, ,
\end{equation}
that has to be studied in the neighborhood of $(a,b)=(0,0)$. In the next appendix we will study the case $c<0$.

The derivative $\widetilde{s}'$ will be zero at $m=0$ and at the values of $m$ satisfying
\begin{equation}
m^6-cm^4+bm^2+a = 0 \, .
\label{derentr}
\end{equation}
Therefore we have to consider the solutions of the last equation as a function of $a$ and $b$, to obtain the phase diagram
in the $(a,b)$ plane. Obviously, because of the symmetry we can restrict the study to the solutions with $m\ge 0$.
By posing $m^2 = x+\frac{c}{3}$ we reduce the equation to the standard form of the third order equation
\begin{equation}
g(x) \equiv x^3 + \left( b-\frac{c^2}{3}\right) x + \left( -\frac{2c^3}{27}+\frac{bc}{3}+a\right) = 0 \, .
\label{redeq}
\end{equation}
Since $m^2 \ge 0$, the only acceptable solutions are those such that
$x \ge -\frac{c}{3}$. Our purpose is to find the maximum in $m$ of Eq. (\ref{entrabc}).

\subsection{$b>\frac{c^2}{3}$}
\label{appcalcp1}
In this case the coefficient of $x$ in Eq. (\ref{redeq}) is positive, so there is only one real solution. Furthermore,
one has $g'(x) = 3x^2 + \left( b-\frac{c^2}{3}\right) >0$, and substituting $x=-\frac{c}{3}$ one obtains
$g\left(-\frac{c}{3}\right)=a$. Therefore the solution is acceptable only for $a \le 0$, and thus only for $a <0$ there
is a solution of Eq. (\ref{derentr}) with $m>0$. On the other hand, since $m=0$ is a local minimum of $\widetilde{s}(m)$
for $a<0$,  the solution with $m>0$ is a maximum. In conclusion, for $b>\frac{c^2}{3}$ the line $a=0$ is a line of
second-order transition (at $m=0$).

\subsection{$b<\frac{c^2}{3}$}
\label{appcalcp2}
This case requires more attention. Now the coefficient of $x$ is $g(x)$ in Eq. (\ref{redeq}) is negative, and from the
general properties, mentioned above, of the solutions of the third order equations, we have that there are three real
solutions of $g(x)=0$ if (after a straightforward calculation)
\begin{equation}
h(a) \equiv 27 a^2 -2(2c^3 - 9 cb)a + b^2(4b-c^2) <0 \, ,
\end{equation}
where we have explicitly indicated the $a$ dependence of the above expression by denoting it as $h(a)$. If, instead,
$h(a) >0$, then $g(x)=0$ has only one real solution. To proceed further, we consider the range in $a$, as a function of $b$,
for which there are three real solutions of $g(x)=0$. To this aim, we compute the roots of $h(a)=0$. They are
\begin{equation}
a_{\pm}= \frac{1}{27} \left[ c(2c^2-9b) \pm 2 \sqrt{(c^2-3b)^3}\right] \, .
\label{rootsh}
\end{equation}
Because of the possible $c$ values we are considering, $a_{\pm}$ are real, and from the sign of the coefficients of $h(a)$,
we have that $a_{\pm}<0$ for $\frac{c^2}{4} < b <\frac{c^2}{3}$, while $a_+>0$ and $a_-<0$ for $b< \frac{c^2}{4}$. We consider
thus the following subcases.
\subsubsection{$ \frac{c^2}{4} < b <\frac{c^2}{3}$}
For the same considerations made above for $b>c^2/3$, the line $a=0$ is still a line of second-order transition (actually,
now we do not have the property that $g'(x)$ is always positive, but since for $a<0$ the value $m=0$ is a local minimum
of $\widetilde{s}(m)$, because of the structure of the latter, see Eq. (\ref{entrabc}), there must be at least a maximum
for $m>0$). But now there is a range of $a$, precisely that with $a_-<a<a_+<0$, for which $g(x)=0$ has three real solutions.
So we now concentrate on $a<0$. For $a<0$ the maximum of $\widetilde{s}(m)$ is always for $m>0$, since $m=0$ is a minimum.
Because of the structure of $\widetilde{s}(m)$ then there is either a maximum or two maxima and a minimum; therefore, of
the three solutions of $g(x)=0$ for $a_-<a<a_+$, either only one is acceptable or all three are acceptable. In addition,
since $g(-\frac{c}{3})=a<0$ and $g'(x)<0$ for $x_-<x<x_+$, with $x_{\pm}= \pm \sqrt{\frac{1}{3}\left(\frac{c^2}{3}-b\right)}$,
and $x_{\pm}>-\frac{c}{3}$, then in all cases the real solutions of $g(x)=0$, either when there is one or when there are
three, are always acceptable. It is possible to see that when $a$ is equal to the mean between $a_-$ and $a_+$, i.e.,
when $a=a^*=\frac{c}{27}(2c^2-9b)$, the two maxima of $\widetilde{s}(m)$ for $m>0$ are equal. In fact, a calculation shows
that for $a=a^*$ the three solutions of Eq. (\ref{derentr}) are
\begin{equation}
0<(m_-^*)^2 =\frac{c}{3}-\sqrt{\frac{c^2}{3}-b}<(m_0^*)^2 = \frac{c}{3}< (m_+^*)^2 =\frac{c}{3}+\sqrt{\frac{c^2}{3}-b} \, ,
\end{equation}
and that
\begin{equation}
\widetilde{s}(m^*_{\pm}) = \left(\frac{c^2}{9}-b\right)^2 >0 \,.
\end{equation}
This finally shows that in this range of $b$ there is, in addition to the second-order transition on the line $a=0$,
a line of first-order transition at $a(b)=a^*$ between two magnetized states (i.e., both of them with $m>0$). This line ends
at a critical point located at $b=\frac{c^2}{3}$ and $a=a^*=-\frac{c^3}{27}$ (where $a_{\pm}=a^*$).

\subsubsection{$\frac{2c^2}{9}<b<\frac{c^2}{4}$}
Now the larger root $a_+$ of $h(a)=0$ (see Eq. (\ref{rootsh})) is positive, so it is possible to have more than one
extremum of $\widetilde{s}(m)$ for $a>0$. Although this is true for any value of $b$ smaller than $\frac{c^2}{4}$, we have
made a further division of the $b$ range, i.e., this one and $b<\frac{2c^2}{9}$, for the reason that will be clear shortly.
Now it is possible in principle to have a first-order transition for $a>0$ between a state with $m=0$ and a magnetized state,
when the entropy $\widetilde{s}(m)$ becomes, at its local maximum, equal to zero from below. To see if this occurs
we adopt the following procedure. At such first-order transition we must have that, for $m$ equal to that of the magnetized
state corresponding to the transition, there is a solution of Eq. (\ref{derentr}) and at the same time
$\widetilde{s}(m)$ vanishes. Thus, we have the following system
\begin{eqnarray}
\label{eqm6m4}
m^6-cm^4+bm^2+a &=&0 \\
m^6 -\frac{4}{3}cm^4+2bm^2+4a &=&0 \, .
\end{eqnarray}
This system can be solved for $m$ and $a$. For example, substracting four times the first equation from the first, one
obtains the equation
\begin{equation}
\label{eqm4}
9m^4-8c m^2 +6b = 0
\end{equation}
Solving this equation one gets $m^2$ as a function of $b$ and $c$; then substituting this solution in the first equation
of the system (\ref{eqm6m4}), one obtains $a$ as a function of $b$ and $c$. At the end, the
relevant solution is given by
\begin{eqnarray}
\label{eqm2xx}
m^2 &=& \frac{1}{9}\left[4c+\sqrt{16c^2-54b}\right] \\
\label{eqaxx}
a &=& \frac{1}{729}\left[ 32 c^3 -162 bc + \frac{1}{2}\left(16c^2-54b\right)^{\frac{3}{2}}\right] \, .
\end{eqnarray}
Let us call $a_t$ this value of $a$. The important point is the following. For the range of $b$ we are considering in this
paragraph, i.e., in particular for $\frac{2c^2}{9}<b<\frac{c^2}{4}$, we have that $a_t<0$. But this means that the entropy
of this extremum of $\widetilde{s}(m)$ reaches the value $0$ when the second-order transition at $m=0$ has already occurred,
and there is necessarily a magnetized state with $\widetilde{s}(m)>0$. Thus, we have to conclude that also in this range
of $b$ we have a second-order transition on the line $a=0$ and a first-order transition between two magnetized states at
$a=a^*=\frac{c}{27}(2c^2-9b)<0$.

\subsubsection{$b<\frac{2c^2}{9}$}
In this range the value $a_t$ obtained in the above procedure is positive. This means that now we have a first-order
transition at the $b$-dependent value $a_t>0$ between an unmagnetized state and a magnetized state. The second-order
transition at $a=0$ becomes a transition between metastable states, and then it does not appear in the phase diagram.

\section{Details of the calculations for $c<0$}
\label{appcalcm}

We now consider $c<0$. Most of the analysis made for the previous case can be used here. The first thing that we have to
note is that in the quadrant where both $a$ and $b$ are positive there cannot be any transition, since the maximum of the entropy
function (\ref{entrabc}) is then at $m=0$, being the coefficients of $m^2$, $m^4$ and $m^6$ all negative.

The whole analysis made in the previous appendix up to the beginning of \ref{appcalcp2} can be repeated unchanged now.
Therefore we conclude that for $b>\frac{c^2}{3}$ the line $a=0$ is a line of second-order transition.

Let us now consider $b<\frac{c^2}{3}$. As in \ref{appcalcp2}, we can obtain the values $a_{\pm}$ given by
Eq. (\ref{rootsh}), between which there can be, potentially, three solution of Eq. (\ref{derentr}). However, we note the following.
Differently from the case $c>0$, now the signs of $a_{\pm}$ are given by
$a_{\pm}>0$ for $\frac{c^2}{4} < b <\frac{c^2}{3}$, while $a_+>0$ and $a_-<0$ for $b< \frac{c^2}{4}$. Since we have seen before
that for $b>0$ there cannot be any maximum of the entropy function (\ref{entrabc}) for $a>0$, we conclude that also for
$\frac{c^2}{4} < b < \frac{c^2}{3}$ there is only a second-order transition at $a=0$.

Going now to $b<\frac{c^2}{4}$ we can proceed as before to obtain the possible values of $m^2$ and $a$, as in
Eqs. (\ref{eqm2xx}) and (\ref{eqaxx}), at a first-order transition. However, we see that the value for $m^2$ is negative,
which obviously is not acceptable. This can be deduced also from Eq. (\ref{eqm4}), that has all positive coefficients
when $b$ is positive and $c$ is negative. On the other hand, when $b$ also is negative, then the solution given
by Eqs. (\ref{eqm2xx}) and (\ref{eqaxx}) is acceptable. It is also possible to see that this value of $a$
is positive, leading to the conclusion, as in the main text, that
there is a line of first-order transitions between an unmagnetized state and a magnetized state, defined by
$a=\frac{1}{729}\left[32c_0^3-162bc_0 +\frac{1}{2}\left(16c_0^2-54b\right)^{\frac{3}{2}}\right]$, $b\le 0$.

\end{document}